\documentclass{emulateapj}
\usepackage{apjfonts}
\usepackage{subfigure}

\newcommand{\der}[2]{\frac{d #1}{d #2}}
\newcommand{\pder}[2]{\frac{\partial #1}{\partial #2}}
\newcommand{\eqref}[1]{Equation (\ref{#1})}
\newcommand{\tabref}[1]{Table~\ref{#1}}
\newcommand{\figref}[1]{Figure~\ref{#1}}
\newcommand{\secref}[1]{\S\ref{#1}}
\newcommand{\logten}[1]{\log_{10}{#1}}
\newcommand{\C}[1]{{}^{#1}{\rm C}}
\newcommand{\0}[1]{{}^{#1}{\rm O}}
\newcommand{\N}[1]{{}^{#1}{\rm N}}
\newcommand{\Ne}[1]{{}^{#1}{\rm Ne}}
\newcommand{\Na}[1]{{}^{#1}{\rm Na}}
\newcommand{\Si}[1]{{}^{#1}{\rm Si}}
\newcommand{\Ni}[1]{{}^{#1}{\rm Ni}}
\newcommand{\Co}[1]{{}^{#1}{\rm Co}}
\newcommand{\Fe}[1]{{}^{#1}{\rm Fe}}
\newcommand{\XC}{X_{12}}
\newcommand{\XNe}{X_{22}}

\newcommand{\XNep}{X_{22}^\prime}
\newcommand{\XNeOp}{X_{22,0}^\prime}
\newcommand{\Ye}{Y_{e}}
\newcommand{\logtenrho}{\mathcal{L}\rho_{\rm DDT}}
\newcommand{\logtenrhoO}{\mathcal{L}\rho_{{\rm DDT},0}}
\shorttitle{SN~Ia dependence on DDT density}

\begin{document}

\submitted{Accepted to the Astrophysical Journal Jul. 6, 2010}
\title{Evaluating Systematic Dependencies of Type Ia Supernovae:\\
The Influence of Deflagration to Detonation Density}

\author{
Aaron P. Jackson\altaffilmark{1},
Alan C. Calder\altaffilmark{1,2},
Dean M. Townsley\altaffilmark{3},
David A. Chamulak\altaffilmark{4,6},\\
Edward F. Brown\altaffilmark{5,6},
and
F.~X. Timmes\altaffilmark{6,7}
}

\altaffiltext{1}{
Department of Physics \& Astronomy,
The State University of New York - Stony Brook, Stony Brook, NY
}
\altaffiltext{2}{
New York Center for Computational Sciences,
The State University of New York - Stony Brook, Stony Brook, NY
}
\altaffiltext{3}{
Department of Physics \& Astronomy,
The University of Alabama, Tuscaloosa, AL
}
\altaffiltext{4}{
Argonne National Laboratory, Argonne, IL
}
\altaffiltext{5}{
Department of Physics \& Astronomy,
Michigan State University, East Lansing, MI
}
\altaffiltext{6}{
The Joint Institute for Nuclear Astrophysics
}
\altaffiltext{7}{
School of Earth and Space Exploration,
Arizona State University, Tempe, AZ
}

\begin{abstract}
We explore the effects of the deflagration to detonation transition
(DDT) density on the production of $\Ni{56}$ in thermonuclear supernova
explosions (type Ia supernovae). Within the DDT paradigm, the transition
density sets the amount of expansion during the deflagration phase of the
explosion and therefore the amount of nuclear statistical equilibrium
(NSE) material produced.  We employ a theoretical framework for a
well-controlled statistical study of two-dimensional simulations of
thermonuclear supernovae with randomized initial conditions that can,
with a particular choice of transition density, produce a similar average
and range of $\Ni{56}$ masses to those inferred from observations.
Within this framework, we utilize a more realistic ``simmered'' white
dwarf progenitor model with a flame model and energetics scheme to
calculate the amount of $\Ni{56}$ and NSE material synthesized for a
suite of simulated explosions in which the transition density is varied
in the range 1--3$\times10^7$~g~cm$^{-3}$.
We find a quadratic dependence of the NSE yield on the log of the
transition density, which is determined by the competition between
plume rise and stellar expansion. By considering the effect of
metallicity on the transition density, we find the NSE yield decreases
by $0.055\pm0.004~M_\odot$ for a $1~Z_\odot$ increase in metallicity
evaluated about solar metallicity. For the same change in metallicity,
this result translates to a $0.067\pm0.004~M_\odot$ decrease in the
$\Ni{56}$ yield, slightly stronger than that due to the variation in
electron fraction from the initial composition. Observations testing the
dependence of the yield on metallicity remain somewhat ambiguous, but
the dependence we find is comparable to that inferred from some studies.
%
\end{abstract}


\section{Introduction}
\label{sec:intro}

Type Ia supernovae are bright stellar explosions that are characterized by
strong P Cygni features in Si and by a lack of hydrogen in their spectra 
(see~\citealt{hillebrandt.niemeyer:type,Fili97} and references therein). 
Observations of type Ia supernovae (serving as distance 
indicators~\citep{phillips:absolute,reispreskirs+96,albrecht_2006_aa}) 
are at present the most powerful and best proved technique for studying
dark energy~\citep{riess.filippenko.ea:observational,
perlmutter.aldering.ea:measurements,KolbReport-of-the-D,Hicketal09b,
Lampeitl10:SDSS}, and, accordingly, there are many observational campaigns
underway striving to gather information about the systematics of these 
events and to better measure the expansion history of the Universe 
(see~\citealt{Kirshner09} and references therein). 

The most widely accepted model for these events is the ``single-degenerate''
scenario, which is the thermonuclear explosion of a white
dwarf composed principally of $\C{12}$ and $\0{16}$ that has
accreted mass from a companion (for a review of explosion models,
see~\citealt{hillebrandt.niemeyer:type,livio2000,Ropke06}). The
peak brightness of the supernova is set by the radioactive decay
of $\Ni{56}$ produced in the explosion to $\Co{56}$ and then to
$\Fe{56}$. The empirical ``brighter is broader'' relation between
the peak brightness of the light curve and the decay time from
its maximum is understood to follow from the fact that both the
luminosity and opacity are set by the mass of $\Ni{56}$ synthesized in the
explosion~\citep{arnett:type,pinto.eastman:physics,Kasen2007On-the-Origin-o}.
Because of the dependence of the light curve on the amount of
radioactive $\Ni{56}$ synthesized during an explosion and the ability
to infer the $\Ni{56}$ yield from observations~\citep{Mazzetal07},
research into modeling thermonuclear supernovae typically focuses
on the production and distribution of $\Ni{56}$ as well as other
nuclides (such as $\Si{28}$) as the measure with which to compare
models to observations.

One-dimensional simulations of the single-degenerate case showed that the
most successful scenario is an initial deflagration (subsonic reaction
front), born in the core of the white dwarf (WD) which at some point
becomes a (supersonic) detonation, i.e. a deflagration-detonation
transition~(DDT,~\citealt{Khokhlov1991Delayed-detonat,HoefKhok96}).
A delayed detonation naturally accounts for the high-velocity Ca
features~\citep{kasen_ca} and the chemical stratification of the ejecta.
While these 1D models are able to reproduce observed features of the
light curve and spectra, much of the physics is missing from these
models.  The presence of fluid instabilities during the deflagration
warranted the development of multidimensional models, allowing a
physically-motivated calculation of the velocity of the burning front
and thus removing a free parameter. By relaxing the symmetry constraints
on the model, buoyancy instabilities are naturally captured leading
to a strong dependence on the initial conditions of the deflagration
that often result in too little expansion of the star by the time DDT
conditions used in previous 1D studies are met~\citep{NiemHillWoos96,
calder.ea:_offset_ignition_1,calder.ea:_offset_ignition_2,
Livne2005On-the-Sensitiv}.  Multidimensional models may reach the
expected amount of expansion prior to the DDT with the choice of
particular ignition conditions and thus retain the desirable features
from one-dimensional models~\citep{gamezo.khokhlov.ea:deflagrations,
PlewCaldLamb04,RoepGiesetal06,Jordan2008Three-Dimension}.

Due to the strong dependence on ignition conditions, multi-dimensional
simulations of the DDT model are able to produce a wide range
peak luminosities (via the production of a range of $\Ni{56}$
yields) consistent with a common explosion mode suggested from
observations~\citep{Mazzetal07}.  Differences in the mass of synthesized
$\Ni{56}$ can follow from properties such as metallicity and central
density of the progenitor and/or differences in the details of the
explosion mechanism such as the density at which the transition from
deflagration to detonation occurs. \cite{timmes.brown.ea:variations} found
that metallicity should affect the $\Ni{56}$ yield based on approximate
lepton number conservation. The metallicity sets the fractional amount
of material in nuclear statistical equilibrium (NSE) that is radioactive
$\Ni{56}$. \cite{bravo10} found a stronger dependence on metallicity
due to a significant amount of $\Ni{56}$ that is synthesized during
incomplete Si-burning.

Observational results to date are consistent with a shallow dependence
of $\Ni{56}$ mass on metallicity but are unable to rule out a trend
entirely~\citep{GallGarnetal05,gallagheretal+08,neilletal+09,howelletal+09}.
Determining the metallicity dependence is challenging not only because
the effect appears to be small, but also due to the difficulty in
measuring accurate metallicities for the parent stellar population
and problems with strong systematic effects associated with the
mass-metallicity relationship within galaxies~\citep{gallagheretal+08}.
This effect is also difficult to decouple from the apparently
stronger effect of the age of the parent stellar population on the mean
brightness of SNe~Ia~\citep{gallagheretal+08,howelletal+09,Krueetal10}.
\citet{howelletal+09} note that the scatter in brightness of this observed
relation is unlikely to be explained by the effect of metallicity. In
general, the source of scatter can be explained by the development of
fluid instabilities during the deflagration phase that contribute to
differing rates of expansion between instances of supernovae. However,
by considering the effect of metallicity on the DDT density, scatter
in the metallicity relation may be enhanced beyond its intrinsic value
inferred from fluid instabilities.

\cite{townetal09} investigated the direct effect of metallicity via
initial neutron excess and found it to have a negligible influence
on the amount of material synthesized to NSE. However, the neutron
excess sets the amount of material in NSE that favors stable Fe-group
elements over radioactive $\Ni{56}$. Therefore, the initial metallicity
directly influences the yield of $\Ni{56}$. In this work, we expand
that study to include the potential indirect effect of metallicity in
the form of the $\Ne{22}$ mass fraction ($\XNe$) through its influence
on the density at which the DDT takes place. To this end, we explore
the effect of varying the transition density, a proxy for varying the
microphysics that determine the conditions for a DDT. 
The conditions under which a DDT occurs are still a subject
of debate.  \citet{NiemWoos97}, \citet{NiemKers97}, and
\citet{KhokOranWhee97} proposed that a necessary condition is the
transition to a distributed burning regime, in which turbulence
disrupts the reaction zone of the flame. More recent numerical
studies~\citep{Panetal08,Woosetal09,Schmetal10} that describe the
conditions for a DDT include dependencies on the turbulent cascade
and the growth of a critical mass of fuel with sufficiently strong
turbulence. For this study, we assume the DDT density to be the density
at which thermonuclear burning is expected to enter the distributed
regime.
This choice links the explosion outcome to the dynamical evolution of
the progenitor density structure during the deflagration phase.
By analyzing the effect of transition density on the NSE yield,
we can later analyze how the details of the microphysics affect the DDT
density. For the purposes of this study, we will consider only the effect
of $\XNe$ on the DDT density. In reality, the $\C{12}$ mass fraction
will also be important in determining the DDT density, but we choose to
leave the exploration of the effect of $\C{12}$ to future work. Many other
possible systematic effects exist that are outlined in \cite{townetal09},
such as the central ignition density~\citep{Krueetal10}, which are all 
held fixed in this study. The interdependence of all of these effects will
be considered in the construction of the full theoretical picture in a
future study.


We describe the details of our model in \secref{sec:progenitor}
and \secref{sec:improvements}, and the properties of our statistical
sample in~\secref{sec:properties}.  We present our findings on the
dependence of transition density on NSE yield in~\secref{sec:ddtdens}. In
\secref{sec:results_nse} and \secref{sec:results_m56}, we assume a
dependence of the transition density on $\Ne{22}$ content and construct
the functional dependence of the $\Ni{56}$ yield on metallicity through
the $\Ne{22}$ content. In \secref{sec:conclusions}, we discuss our
conclusions and future work.

\section{Parameterized Realistic White Dwarf Progenitor}
\label{sec:progenitor}

In order to include relevant processes in explosion models, we first
estimate the compositional profile of the progenitor white dwarf just
prior to the birth of the flame.  We begin by estimating the compositional
profile resulting from the evolution of the post-main-sequence star that
later becomes a white dwarf. Recall that the initial metallicity of
the star is, by mass, almost entirely in the form of CNO. As a result
of the C-N-O cycle, these all end up in helium layers as $\N{14}$,
the target of the slowest proton-capture in the cycle.  Subsequent
reactions during helium burning convert $\N{14}$ into $\Ne{22}$;
therefore, $\XNe$ is proportional to the initial metallicity of the
star~\citep{timmes.brown.ea:variations}. The composition of the inner
portion of the star ($\approx~0.3-0.4~M_\odot$) is set during core helium
burning resulting in a reduced carbon mass fraction with respect to that
of the outer layers, which is set by shell burning on the asymptotic
giant branch (see~\citealt{Straetal03}, and references therein).

\begin{figure}[tbh]
\plotone{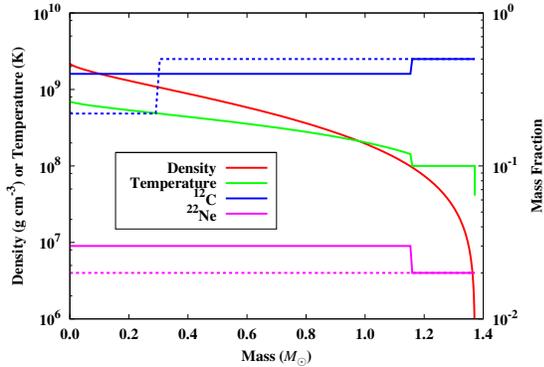}
\caption{\label{fig:prog}Composition, density and thermal 
profiles of the progenitor white dwarf star used in the simulations for 
this study (solid lines).  The compositional profile of the progenitor 
prior to the simmering phase is also shown (dashed lines).
}
\end{figure}

At some point after the white dwarf has formed, it begins to accrete material 
from its companion. As the mass of the accreting white dwarf approaches 
the Chandrasekhar mass limit, the core temperature and density increase such 
that carbon begins to fuse.  The energy released by carbon burning drives 
convection in the core.  The convective carbon burning (``simmering'') phase 
lasts approximately 1000 years before the central temperature is high enough 
to spark a thermonuclear flame~\citep{WoosWunsKuhl04}. During the simmering 
phase, carbon is consumed from the convective core. Concurrently, though, the 
convective zone grows with increasing central temperature, pulling in 
relatively carbon-rich material from the outer layer. Thus, the net effect is 
to increase the carbon abundance in the convective region.  We show the growth
of the convective zone in \figref{fig:prog}, where dashed lines show the 
compositional profile prior to the simmering phase and the corresponding solid
lines show the compositional profile at the start of our simulations of the 
explosion.

For our WD models, we consider a parameterized 3-species compositional
structure consisting of $\C{12}$, $\0{16}$, and $\Ne{22}$, which is
sufficient to capture the nuclear burning rates.  Since $\Ne{22}$ is the
only element in our parameterization that has more neutrons than protons,
the neutron excess from simmering is accounted for by a parameterized
$\Ne{22}$ mass fraction
\begin{equation}
\label{eq:x22param}
\XNep = \XNe + \Delta\XNe(\Delta\Ye) {\rm ,}
\end{equation}
where $\XNe$ is proportional to the initial metallicity coming from helium
burning~\citep{timmes.brown.ea:variations} and $\Delta\XNe(\Delta\Ye)$
represents the effective enhancement of $\Ne{22}$ in the core following
from the change in electron fraction during convective carbon burning. The
electron fraction ($\Ye$) is related to $\XNe$ by
\begin{equation}\label{eq:ye}
\Ye = \frac{10}{22} \XNe + \frac{1}{2}(1-\XNe)
\end{equation}
such that a change in $\XNe$ constitutes a change in $\Ye$.  The prime
on $\XNep$ in \eqref{eq:x22param} indicates inclusion of effects of
neutronization during carbon burning and this convention applies to
expressions below.  We choose the composition of the WD at the end of the
simmering phase to consist of $\XC = 0.4$ and $\XNep = 0.03$ in the core
and $\XC = 0.5$ and $\XNep = 0.02$ in the outer layer. Note that $\XNe =
\XNep$ in the outer layer because neutronization due to carbon burning
only occurs in the convective core.

For comparison, a compositional profile can be estimated for the
WD prior to the onset of carbon burning. For simplicity, consider
the production of one neutron for every 6 $\C{12}$ burned (i.e.,
$d\Ye/dY_{12} \approx 1/6$, where $Y_{12}$ is the molar abundance of
$\C{12}$)~\citep{Chametal08,PiroBild08}.  For $\Delta\XNe = 0.01$ in our
progenitor model, this constitutes burning $0.04~M_\odot$ of carbon during
simmering. Assuming that prior to simmering, the core mass is $\approx
0.3~M_\odot$~\citep{Straetal03} and the outer layer consists of $\XC =
0.5$ material, we can account for all $\C{12}$ and conserve the total
mass of the WD to estimate $\XC \approx 0.2$ in the carbon-reduced core
of the WD prior to simmering as shown in \figref{fig:prog}.

\tabref{tab:prog} shows the composition of the progenitor white dwarf
before and after the simmering phase using the parameterized $\Ne{22}$
mass fraction as well as the core mass. Note that throughout this study,
we choose to neglect any variation of the amount of neutronization during
simmering due to the initial $Z$.  Accordingly, $\Delta\XNe(\Delta\Ye)$
is treated as a constant and, therefore, ${\rm d}\XNep = {\rm d} \XNe$
and derivatives involving the true $\Ne{22}$ mass fraction proportional
to metallicity, $\XNe$, are equivalent to derivatives involving the
parameterized $\Ne{22}$ mass fraction, $\XNep$.

\begin{table}
\caption{\label{tab:prog}Composition of the core and outer layer of the 
progenitor white dwarf.
}
\begin{center}
\begin{tabular}{l||c|c|c|c||c|c|c}
& \multicolumn{4}{c||}{Core} & \multicolumn{3}{c}{Outer Layer} \\
\hline
& $\XC$ & $X_{16}$ & $\XNep$ & $M_{\rm c}$ ($M_\odot$) & 
  $\XC$ & $X_{16}$ & $\XNep$ \\ 
\hline
Pre-Simmering & 0.22 & 0.76 & 0.02 & 0.30 & 0.50 & 0.48 & 0.02 \\
Pre-Deflagration & 0.40 & 0.57 & 0.03 & 1.16 & 0.50 & 0.48 & 0.02 \\
\hline
\end{tabular}
\end{center}
\end{table}

Because we consider the enhancement of the laminar flame speed by
$\Ne{22}$ \citep{chamulak+07}, we need to consider the effects of our
parametrization of the $\Ne{22}$ content with care. In actuality, the
neutronization during carbon simmering produces $\C{13}$, $\Na{23}$, and
$\Ne{20}$ and not $\Ne{22}$~\citep{Chametal08}. A priori, this difference
could alter the nuclear burning rates and hence the laminar flame speed
and width.  \tabref{tab:ne22ne23} shows the laminar flame speeds and
widths for the same $\Ye$ exchanging $\Delta\XNe$ for carbon-simmering ash
for a composition of 40\% $\C{12}$, 2.0\% $\Ne{22}$, 55.5\% $\0{16}$,
0.6\% $\C{13}$, 0.9\% $\Ne{20}$, and 1.0\% $\Na{23}$ by mass using
the same method as described by \cite{chamulak+07}. The flame speeds
and widths calculated using the parameterized $\Ne{22}$ mass fraction
(3\% by mass) are denoted with primes.  For high densities ($\gtrsim
2.5\times10^8$~g~cm$^{-3}$) in which the flame speed is not dominated
by the buoyancy-driven Rayleigh-Taylor instability, the difference is
$\lesssim 5\%$; therefore, our parametrization of the neutronization via
$\XNe$ accurately captures the corresponding enhancement of the laminar
flame speed in this regime.

\begin{table}
\caption{\label{tab:ne22ne23}Flame speeds and widths changing $\Ne{22}$ to 
carbon-simmering ashes holding $\Ye = 0.498636$ fixed and using $\XC = 0.4$.
Primed quantities parameterize the effects of neutronization
during carbon burning as additional $\Ne{22}$.
}
\begin{center}
\begin{tabular}{c||c|c|c}
$\rho_9$ (g cm$^{-3}$) & $s^\prime$ (km s$^{-1}$) & $s$ (km s$^{-1}$) & 
Diff. (\%) \\
\hline
0.1 &  0.926 &  1.012 & 8.5 \\
0.2 &  4.194 &  4.570 & 8.2 \\
0.3 & 11.372 & 12.106 & 6.1 \\
0.4 & 18.785 & 19.466 & 3.5 \\
0.5 & 24.352 & 25.057 & 2.8 \\
0.6 & 29.162 & 29.916 & 2.5 \\
0.7 & 33.527 & 34.322 & 2.3 \\
0.8 & 37.571 & 38.401 & 2.2 \\
0.9 & 41.364 & 42.228 & 2.0 \\
1.0 & 44.978 & 45.871 & 1.9 \\
\hline
\hline
$\rho_9$ (g cm$^{-3}$) & $\delta^\prime$ (cm) & $\delta$ (cm) & Diff. (\%) \\
\hline
0.1 & $1.4018\times10^2$ & $1.3829\times10^2$ & -1.4 \\
0.2 & $2.7439\times10^3$ & $2.4097\times10^3$ & -13.9 \\
0.3 & $8.5650\times10^4$ & $8.1055\times10^4$ & -5.7 \\
0.4 & $4.6600\times10^4$ & $4.9033\times10^4$ & -5.0 \\
0.5 & $3.5373\times10^4$ & $3.4319\times10^4$ & -3.1 \\
0.6 & $2.6457\times10^4$ & $2.5756\times10^4$ & -2.7 \\
0.7 & $2.0980\times10^4$ & $2.0437\times10^4$ & -2.7 \\
0.8 & $1.7167\times10^4$ & $1.6820\times10^4$ & -2.1 \\
0.9 & $1.4557\times10^4$ & $1.4199\times10^4$ & -2.5 \\
1.0 & $1.2562\times10^4$ & $1.2251\times10^4$ & -2.5 \\
\hline
\end{tabular}
\end{center}
\end{table}

While the central temperature and central density of the progenitor just prior 
to the birth of the flame are primarily set by the accretion history of the 
white dwarf, which varies and is largely unknown, we choose a fiducial central 
density of $\rho_c = 2.2\times10^9$ g cm$^{-3}$ and central temperature of 
$T_c = 7\times10^8$ K. We construct isentropic profiles of density and 
temperature in the (convective) core and isothermal profiles in the (thermally 
conductive) outer layer while maintaining hydrostatic equilibrium. We choose a 
fiducial isothermal temperature of the outer layer to be $T_{\rm iso} =
10^8$~K. The total mass of the WD progenitor model is $1.37~M_\odot$.

Due to the difference in composition between the core and outer layer, 
requiring a neutral buoyant condition at this interface no longer reduces to a 
continuous temperature as it does for a compositionally uniform white dwarf
\citep{PiroChan08}. In reality, we might expect some convective 
overshoot and mixing between the core and the outer layer, but because the 
physics of convective overshoot are complex and not well understood, the width
of the transition region is unknown. For simplicity, we assume no mixing
region. The composition, density, and thermal profiles of the progenitor 
used for this study immediately prior to deflagration are the solid lines
shown in \figref{fig:prog}.

\section{Simulation Methods}
\label{sec:improvements}

We use a customized version of the FLASH Eulerian compressible adaptive-mesh
hydrodynamics code~\citep{Fryxetal00,calder.fryxell.ea:on}. Modifications
to the code include a nuclear burning model, composition-dependent laminar 
flame speeds, particular mesh refinement criteria, and instructions for
determining the conditions for a DDT~\citep[for details, see][]{Caldetal07,
Townetal07, townetal09}. In particular, the adaptive mesh refinement (AMR)
capability of the code was utilized to achieve particular resolutions for
burning fronts (4~km) and the initial hydrostatic star (16~km) for which
the solution has converged~\citep{townetal09}.


The burning model is used for both subsonic (deflagration) and supersonic 
(detonation) burning fronts. The laminar flame width for densities and 
compositions characteristic of a massive C-O WD is unresolved on the scale 
of our grid (4~km, \citealt{chamulak+07}); therefore, we use an artificially
thickened flame represented by the advection-reaction-diffusion
equation~\citep{Khok95,VladWeirRyzh06} to which our nuclear energetics
scheme is coupled. Appropriate measures have been taken to ensure this
coupling is acoustically quiet and stable such that the buoyant instability
of the burning front is accurately captured~\citep{Townetal07}. The 
shock-capturing capabilities of FLASH naturally handle the propagation
of detonation fronts given an accurate nuclear energetics 
scheme~\citep{Meaketal09}. We note that the only common 
components between our code and 
that of \citet{Plew07} are those publicly available as components of FLASH, 
which excludes all components treating the nuclear burning; differences are 
discussed in \citet{Townetal07}.

In this section we discuss an improvement to our burning model and an 
improvement to the method by which a transition from deflagration to detonation
is made. The changes to the burning model reflect recent work to better match
steady-state detonation structures that are partially resolved on the grid.
This is important for obtaining accurate particle tracks for post-processing
nucleosynthetic yields. Additionally, changes to the DDT method were
necessary to ensure consistency between individual simulations using different
DDT densities and/or different initial configurations of the flame.

\subsection{Improved Burning Model}
\label{sec:burning}

For the calculations presented here we utilize the latest revision of a
parameterized three-stage model for the nuclear burning~\citep{Caldetal07,
Townetal07,Meaketal09,SeitTownetal09,townetal09}. The details
of this latest version will be published separately~(Townsley et al., 
in prep.) along with extensive comparisons to nuclear network calculations 
of steady-state detonations, but we summarize the major changes here.  This
work represents the first time a burning model which correctly reproduces 
the nuclear statistical quasi-equilibrium (NSQE, \citealt{Khok89}) to NSE
transition time scales and length scales during incomplete 
silicon burning has ever been used in a multi-dimensional SN~Ia calculation,
as validated by comparison to steady-state detonation structures calculated
out to the pathological point
with the ZND model \citep[see e.g.][]{FickDavi79}.  Accurate reproduction of
this low-density burning regime is essential because a significant portion
of the $^{56}$Ni is produced in incomplete burning~\citep{bravo10} so that 
the overall $^{56}$Ni yield is determined by the details of how this 
low-density cutoff of $^{56}$Ni production occurs.

Obtaining a satisfactory reproduction of ZND detonation structures involved
two main changes to the 3-stage burning model.  First, we found that the
progress variable representing the NSQE to NSE transition, $\phi_{qn}$, 
which also gives the mass fraction of Fe-group (NSE) material,
did not match the time and space structure of this transition in steady 
state detonations calculated with a full nuclear network.  The kinetics for
this stage, first proposed by \citet{Khokhlov1991Delayed-detonat} and adopted 
in \citet{Caldetal07}, are given by the simple form $d\phi_{qn}/dt =
(1-\phi_{qn})/\tau_{\rm NSE}$ where $\tau_{\rm NSE}$ is a calibrated timescale
that depends mainly on temperature.  We have found, however that a far more
appropriate match to steady-state detonations at densities important for
incomplete silicon burning, $\rho \lesssim 10^7$~g~cm$^{-3}$, is obtained
with the alternative form $d\phi_{qn}/dt=(1-\phi_{qn})^2/\tau_{\rm NSE}$.
This necessitates a recalibration of $\tau_{\rm NSE}$, since it now plays a
different role, but it is still sufficient for it to depend only on
temperature. Although this change in derivative significantly
improves the match between the parameterized burning and the ZND structure at
the densities of interest, it is still not exact at all densities.  It is
therefore unclear if this form is indicative of some underlying physical
process, and whether or not it is specific to detonations.  There are several
relaxation processes proceeding simultaneously, so that it is non-trivial to
quantify separate contributions.  This will be investigated in more detail in
future work on post-processing nucleosynthesis.

The second major burning model change was motivated by needing to match the
thermodynamic, i.e.  $T$, $\rho$, profiles at densities at which the portion
of the detonation structure representing the NSQE to NSE transition is
resolved on the grid.  Our previous treatment released all of the
nuclear binding energy change to the NSE state by the end of the second
stage, leaving the third stage NSQE to NSE transition energetically inert.
This lack of an energy release on the NSE timescale leads to an incorrect
progression of the density fall-off behind the shock front in the detonation.
The abbreviated energy release leads to an under-prediction of the temperature
immediately behind the unresolved earlier burning stages in a propagating
detonation.  A very good match of thermodynamic profiles to the full-network
steady state detonation was obtained by tying the completion of the energy
release to $\phi_{qn}$, so that energy release occurs in three distinct
stages.  Note that the previous change involving the kinetics used for
$\phi_{qn}$ is also an important contributor to the realism of the
thermodynamic profiles obtained.

Finally, although detailed nucleosynthesis based on post-processing tracer
particle histories will be published in future work (Townsley et al., in
prep.), we have performed two important tests for the burning model used in
this work.  The composition profiles obtained from post-processed histories
for hydrodynamic simulations of steady-state detonations match the
steady-state structure calculated via the ZND method with a large network
with remarkable accuracy.  Additionally, a preliminary version of the
post-processing under development has been applied to the calculations
presented here, and we have found good agreement between $\phi_{qn}$ and the
the fraction of material in the form of Fe-group nuclides found from
post-processing on both an overall basis and in ejection velocity bins.
Overall, the burning model changes led to a modest but significant (around
0.2$M_\odot$) increase in the overall Fe-group yields for the same explosion.
The yields found here are not quite this much higher than similar cases from
\citet{townetal09} because the progenitors used here have a lower central
carbon fraction.

\subsection{Improved Detonation Ignition Conditions}
\label{sec:autodet}

\begin{figure}[tbh]
\plotone{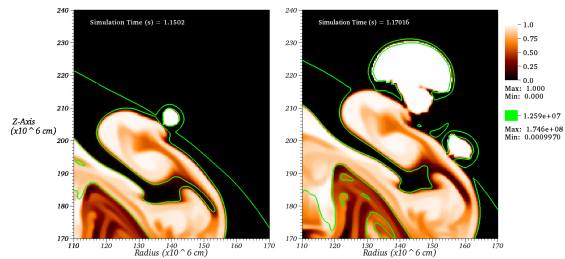}
\caption{\label{fig:autodet}Snapshots of realization 2 just after one (left)
and then another (right) detonation was ignited at the specified transition 
density of $1.26 \times 10^7$ g cm$^{-3}$ (green contour). The reaction 
progress variable representing carbon burning is in color. The detonating 
regions are the rapidly expanding regions of completely burned carbon ahead
of the plume.
}
\end{figure}

In order to study the systematic effects associated with changing the 
DDT density, we need to minimize any systematics in our method of 
starting a detonation. Previously, in \cite{townetal09}, we visualized
the simulation data from the deflagration phase and plotted a density 
contour at the DDT density.  When the flame reached $\approx 64$~km away
from the contour, we picked a computational cell half-way between the 
flame front and the DDT density contour to place a detonation ignition
point with a radius of 8~km. The placement and size
of the detonation point was chosen to be as close to the flame front as
possible while still allowing the detonation point to develop into a
self-sustained, stable detonation front. If the detonation point is
placed too close to the flame front, then the flame will interact with
the detonation point before it develops into a self-sustained
detonation. Comparisons of simulations from identical initial
conditions with 8 and 12~km ignition radii finds the NSE yield differs
by $\lesssim 0.5\%$ throughout the evolution. This results indicates
that the total yield is insensitive to the choice of detonation ignition
radius for radii less than the characteristic size of a rising plume (as
can be seen in \figref{fig:autodet}). We adopt 12~km for the detonation
ignition radius in our study as it produces more robust detonations at
low density.

To ensure that we do not introduce unintended systematic effects in this
study, we improve our method of detonation ignition point placement over
the previous ``by hand'' method by precisely defining the criteria for
a DDT that is used in an algorithm. Parameters in this
algorithm are chosen to be consistent with \citet{townetal09}. Once
the flame front reaches the specified DDT density in a simulation,
a detonation is ignited 32~km radially outward away from the flame
front. Here the reaction-diffusion (RD) front is defined by the variable
representing progress of the subsonic burning wave, $\phi_{\rm RD}$,
with the leading edge defined as the region between the values 0.001 and
0.01 of this variable. During the deflagration phase of a simulation,
$\phi_{\rm RD}$ is equivalent to the carbon-burning reaction progress
variable, $\phi_{fa}$.  At the leading edge of the RD front, very little
carbon has burned and the local density is approximately equal to the
unburned density. This provides a definition of the DDT density that is
much more precise and accurate. We chose these criteria that ignite the
detonation ahead of the RD front to avoid any issues with the detonation
ignition point overlapping with the artificially thickened flame. If
we were to choose criteria that would initiate a detonation inside a
thickened flame, the detonation structure would need to be altered in
some way to be consistent with the artificial nature of that region.

Our detonation ignition conditions also restrict detonation ignition points 
to be at least 200~km away from each other. This choice ensures that each 
rising plume starts 2--3 detonations, which is consistent with
\cite{townetal09}. In the case that multiple points
within 200~km meet the detonation ignition conditions, the point furthest from
the center of the star is preferred ensuring the ignition of detonations on
plume ``tops''. In reality, the location of the spontaneous
detonation points is not well known and is the subject of active
research. For instance, \citet{Roepetal07} argue that a spontaneous detonation
is triggered by the extreme turbulence found in the roiling fuel underneath
the plume caps. Under the assumption that the DDT occurs when the flame
reaches the low density for distributed burning, we place ignition points
on the tops of the rising plumes. Future studies, however, 
will explore other physically motivated detonation methods in which the
location of the detonation ignition is not necessarily specified relative
to a plume, but rather determined by the local turbulence field, composition,
density, and temperature~\citep{Roepetal07,Woosetal09}.
\figref{fig:autodet} shows the placement of first one detonation point, and
then another more than 200~km away from the first, both at the specified
DDT density.

Each detonation ignition point is defined by the profile of the reaction 
progress variable representing carbon burning, $\phi_{fa}$. Within the 
detonation ignition radius $\phi_{fa} = 1$, and, because we are igniting in 
fuel, $\phi_{fa} = 0$ outside this region. The subsequent energy release
from this change in $\phi_{fa}$ drives a shock strong enough to ignite
the surrounding unburned material. This top-hat profile is a simple 
approach to igniting a detonation and has some drawbacks. At densities below 
$\approx10^7$~g~cm$^{-3}$, this detonation ignition method does not produce a 
sufficiently strong shock to burn the fuel. The improvements to our DDT
method place the detonation point radially outward from the specified DDT
density; therefore, the density within the detonation point is actually
somewhat less than that specified by the DDT density. This differs from the
method described in \citet{townetal09} in that the detonation point was
placed at a density that was somewhat higher than the specified DDT
density. Igniting a detonation specifying a DDT density below 
$10^{7.1}$~g~cm$^{-3}$ is impossible using a top-hat profile because the
detonation ignition point is actually placed $\approx10^7$~g~cm$^{-3}$. In
future works, we will explore the use of a gradient in $\phi_{fa}$
motivated by \citet{Seitetal09} to describe the detonation ignition point
profile which should result in the formation of stronger shocks that will
burn the surrounding low-density fuel.

\section{Properties of Statistical Sample and Method}
\label{sec:properties}

To study the systematic effect of transition density on the $\Ni{56}$ yield 
synthesized during a simulated explosion, we utilize the statistical framework 
developed in \cite{townetal09}. In order to compare results between this study
and \cite{townetal09}, we use the same sample population using the same initial 
seed.  The initial seed defines the starting point to a stream
of random numbers used to characterize our sample population of thermonuclear
supernovae as described in \citet{townetal09}. In that study, we found that 
the sample dispersion in the estimated NSE yield does not asymptote until more 
than 10 realizations. Accordingly, our sample is made up of 30 realizations.

During the early part of the deflagration phase ($\approx
0.1$~s), the flame is most affected by convective motion in the core of
the WD. Because the velocity field in the core and the number of
ignition points are largely uncertain, we choose to characterize the
initial flame surface using spherical harmonics, each with a random
coefficient picked sequentially from the initial seed. Each realization
is defined with a unique perturbation on the initial spherically
symmetric flame surface using
\begin{equation}
\label{eq:pert}
r\left(\theta\right) = r_0 + 
a_0 \sum_{l=l_{\rm min}}^{l_{\rm max}} A_l Y_l^0\left(\theta\right) {\rm ,}
\end{equation}
where $r_0$ is the radius of flame surface, $a_0$ is the amplitude of
the perturbations, $A_l$ is a randomly-chosen coefficient corresponding
to the spherical harmonic $Y_l^0$, and $l_{\rm min}$ and $l_{\rm max}$
set the range of spherical modes used to perturb the flame surface. This
method serves to initialize Rayleigh-Taylor unstable plumes of random
relative strengths, which we would expect from various distributions of
ignition points and varying strengths of the convective velocity field
found in the real population of progenitors. In this study,
$r_0 = 150$~km, $a_0 = 30$~km, $l_{\rm min} = 12$, and $l_{\rm max} =
16$. The choices of these parameters are motivated by the resolution of
our study and the desire to obtain reasonable $\Ni{56}$ yields inferred
from observations. These choices are discussed in further detail in
\cite{townetal09}.

We choose to analyze the dependence on transition density by choosing 5
different transition densities equidistant in log space at $\logtenrho
= \{7.1, 7.2, 7.3, 7.4, 7.5\}$ in cgs units, where $\logtenrho =
\logten{\rho_{\rm DDT}}$. The range of DDT densities chosen
for this study are motivated both by previous work and computational
challenges. Below $\logtenrho = 7.1$, a more realistic detonation
structure is needed to successfully launch a detonation
wave as described in the previous section and in \citet{Seitetal09}. While
recent studies have suggested a wider range of DDT densities are
possible~\citep{Panetal08,Schmetal10}, we suggest that trends resulting
from varying the DDT density are captured with a maximum $\logtenrho
= 7.5$.

A simulation is performed for each of the 30 realizations in our sample
at each DDT density for a total of 150 simulations. We choose not
to explore DDT densities below this specified range for computational
reasons. Because of the approximate power-law density profile of the WD,
DDT densities were chosen in log space because these densities correspond
to relatively evenly spaced radial coordinates of the WD.  The amount of
$\Ni{56}$ synthesized in the explosion principally depends on the amount
of expansion during the deflagration phase. Therefore, spatially separated
detonation ignition conditions will allow the amount of expansion to vary
a non-negligible amount and we can more easily analyze the dependence
of the yield on DDT density.

\begin{figure}[tbh]
\plotone{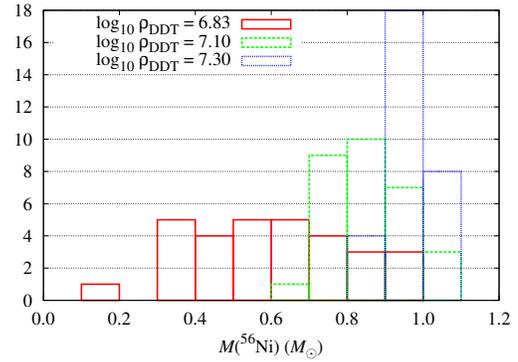}
\caption{\label{fig:histogram}The distributions of the $\Ni{56}$
yield for three different transition densities $\rho_7$
= 0.76 (solid), 1.26 (dashed), and 2.00 (dotted). Note that the variance 
increases with decreasing transition density. The distribution corresponding
to $\rho_7=0.76$ is calculated by extrapolating the dependence of DDT density
as described in \secref{sec:ddtdens} such that the mean 
$M(\Ni{56}) \simeq 0.60~M_\odot$ consistent with observations.
}
\end{figure}  

We find that extrapolating our results to a DDT density of $\logtenrho =
6.83$ yields an estimated average $\Ni{56}$ yield of $\simeq 0.60~M_\odot$
with a standard deviation of $0.21~M_\odot$ (from the fits described
below and listed in \tabref{tab:nsefit}) that is consistent with
observations~\citep{howelletal+09}. Because the actual DDT density is
unknown and the subject of ongoing research~\citep{Panetal08,Aspdetal08,
Aspdetal10,Woosetal09,Schmetal10}, we are free to choose this value of the
DDT density as the fiducial DDT density, $\logtenrhoO = 6.83$. This choice
is relevant for analysis and comparison to other works as discussed in
\secref{sec:results}.  The distribution of $\Ni{56}$ material synthesized
during the explosion is shown in \figref{fig:histogram} for different
transition densities at $\logtenrho = 6.83,~7.10,~{\rm and}~7.30$.

\begin{figure}[tbh]
\plotone{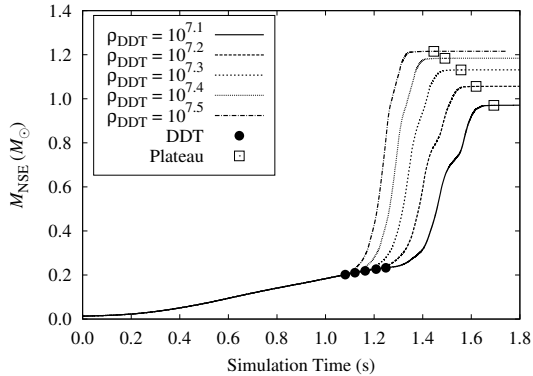}
\caption{\label{fig:nse_evol}The NSE yields for realization 2 at
each transition density used in this study.  The closed circles indicate the
time the flame first reaches the DDT density. The open squares show the time 
the NSE yield has plateaued as defined by \eqref{eq:critNSE}.
}
\end{figure}

The NSE mass, $M_{\rm NSE}$, is defined as $\int \phi_{qn} \rho {\rm d}V$
integrated over the star.  We determine the NSE yield by running the simulation 
until $M_{\rm NSE}$ is no longer increasing as a function of simulation time.
This condition is defined as 
\begin{equation}
\label{eq:critNSE}
\der{M_{\rm NSE}}{t} < 0.01~\frac{M_\odot}{s}{\rm .}
\end{equation}
Because we estimate the $\Ni{56}$ yield as a fraction of the NSE yield, we
consider the $\Ni{56}$ yield to have plateaued when the NSE yield has 
plateaued. 
Considering additional $\Ni{56}$ from incomplete Si-burning in NSQE material
and the efficient capture of excess neutrons onto Fe-group elements changes
the final $\Ni{56}$ estimate by $\lesssim 1\%$. Therefore, to good
approximation, the $\Ni{56}$ yield is a fraction of the NSE yield.
\figref{fig:nse_evol} shows the evolution of the NSE yield as a function of 
simulation time showing the DDT time and the NSE yield plateau time for each 
DDT density for realization 2. Discernible from this figure, there is a clear
dependence of the NSE yield on transition density.

\section{Results}
\label{sec:results}

Within the DDT paradigm for thermonuclear supernovae, the
conditions under which a transition occurs are still not completely
understood~\citep{Aspdetal08,Aspdetal10}.  Generally, though, a
hypothesized transition occurs when the flame enters the distributed
burning regime, which occurs when the flame speed equals the turbulent
velocity at the scale of the flame width assuming a Kolmogorov turbulent
cascade~\citep{NiemWoos97}. Recent work has placed more
stringent requirements on the DDT~\citep{Panetal08,Woosetal09,Schmetal10}
and fundamental questions concerning deflagration in the limit of
disruptive turbulence remain~\citep{PoluOran10}. Regardless of the
actual DDT mechanism, it is likely that the DDT conditions will depend
on composition because both the width and burning rate of the flame
depend on the abundances $\C{12}$ and $\Ne{22}$. For the current study,
we assume that a DDT will occur at a unique density given a particular
composition, regardless of the microphysics involved. This assumption is
reasonable given that the characteristics of the flame depend strongly
on density.  Using this assumption, we can delay the analysis of the
particular microphysics that lead to a specific transition density and
analyze the dependence of the amount of material synthesized to NSE
during the explosion on the transition density. Therefore, we can think
of each transition density as a proxy for changing the composition that
determines the conditions for a DDT via the appropriate microphysics.

As discussed in detail in \citet{townetal09}, many systematic effects exist
that influence the outcome of a SN. In that study, the direct effect of 
$\Ne{22}$ was explored and found to have a negligible influence on the
NSE yield. Therefore, we do not vary the initial $\Ne{22}$ mass fraction,
but rather study the effect of varying the DDT density with the expectation 
that the NSE yield will be influenced indirectly by $\Ne{22}$ and $\C{12}$ 
abundances through the DDT density. We focus on the indirect effect of the 
$\Ne{22}$ abundance on the NSE yield, following up previous work in 
\citet{townetal09} and neglect the effect of varying the carbon abundance.
Additionally, we neglect effects due to the central ignition density, 
compositional and thermodynamic WD structure, and the total WD mass. The 
effects due to these variables will be studied in turn in future works with
the goal of addressing interdependencies once individual effects are better
understood.

\subsection{Dependence on Transition Density}
\label{sec:ddtdens}

\begin{figure}[tbh]
\plotone{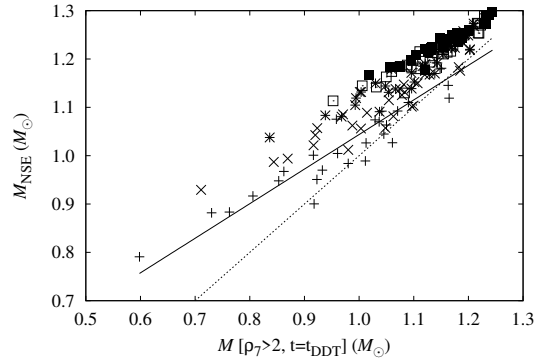}
\caption{\label{fig:Mrho}Mass burned to NSE as compared to the mass above a 
density of $2 \times 10^7$~g~cm$^{-3}$ at the first DDT time for each 
realization in our sample.  The different shapes plotted (+, $\times$, $\ast$,
$\Box$, $\blacksquare$) correspond to different transition densities 
($10^{7.1}$, $10^{7.2}$, $10^{7.3}$, $10^{7.4}$, $10^{7.5}$), respectively.
The solid line shows the linear fit to yields produced with a DDT density of 
$10^{7.1}$~g~cm$^{-3}$ most closely matching the DDT density used in 
\cite{townetal09}. The dashed line shows a 1:1 correlation between the NSE 
yield and mass above $2 \times 10^7$~g~cm$^{-3}$. The lowest two DDT
densities are less than the density threshold $2 \times 10^7$~g~cm$^{-3}$
and thus show more scatter about the linear relation due to increased
dependence on the plume morphology.}
\end{figure}

\begin{figure*}[t!]
\begin{center}
\plottwo{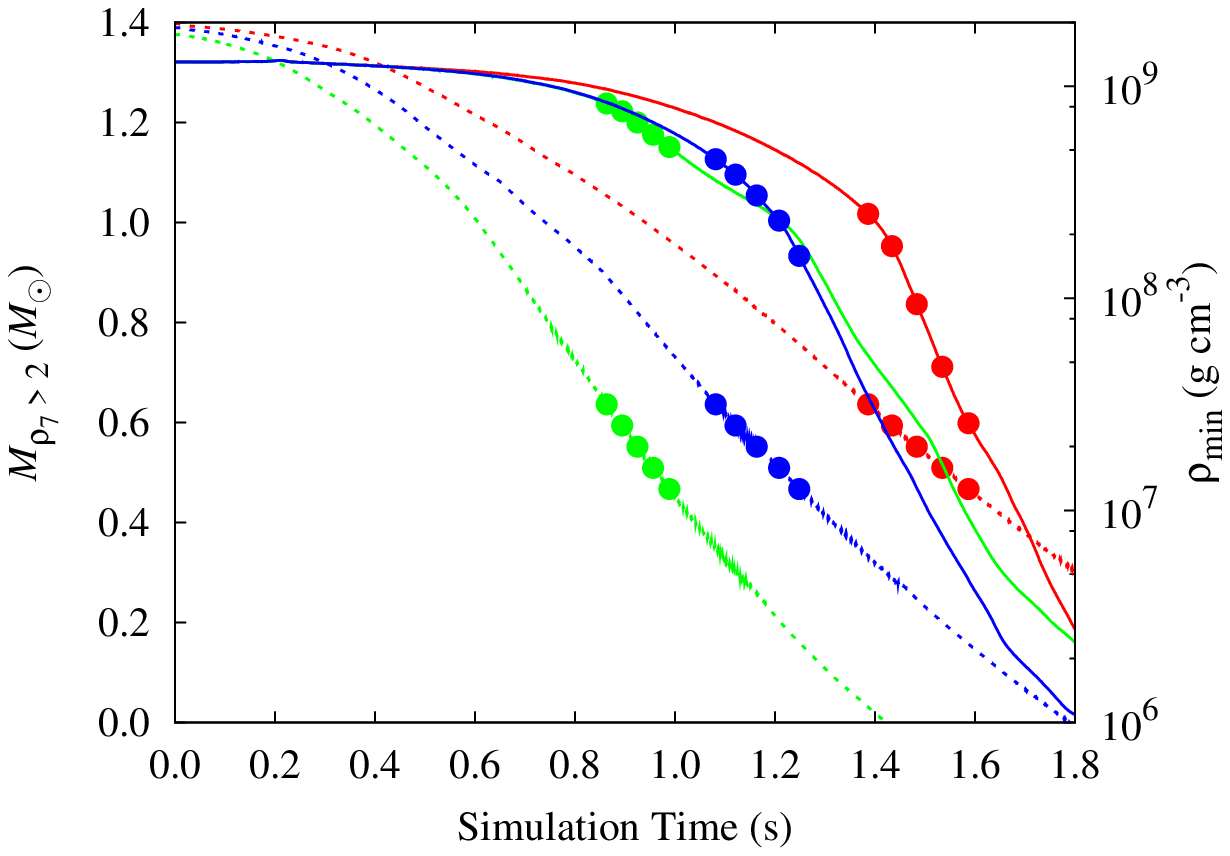}{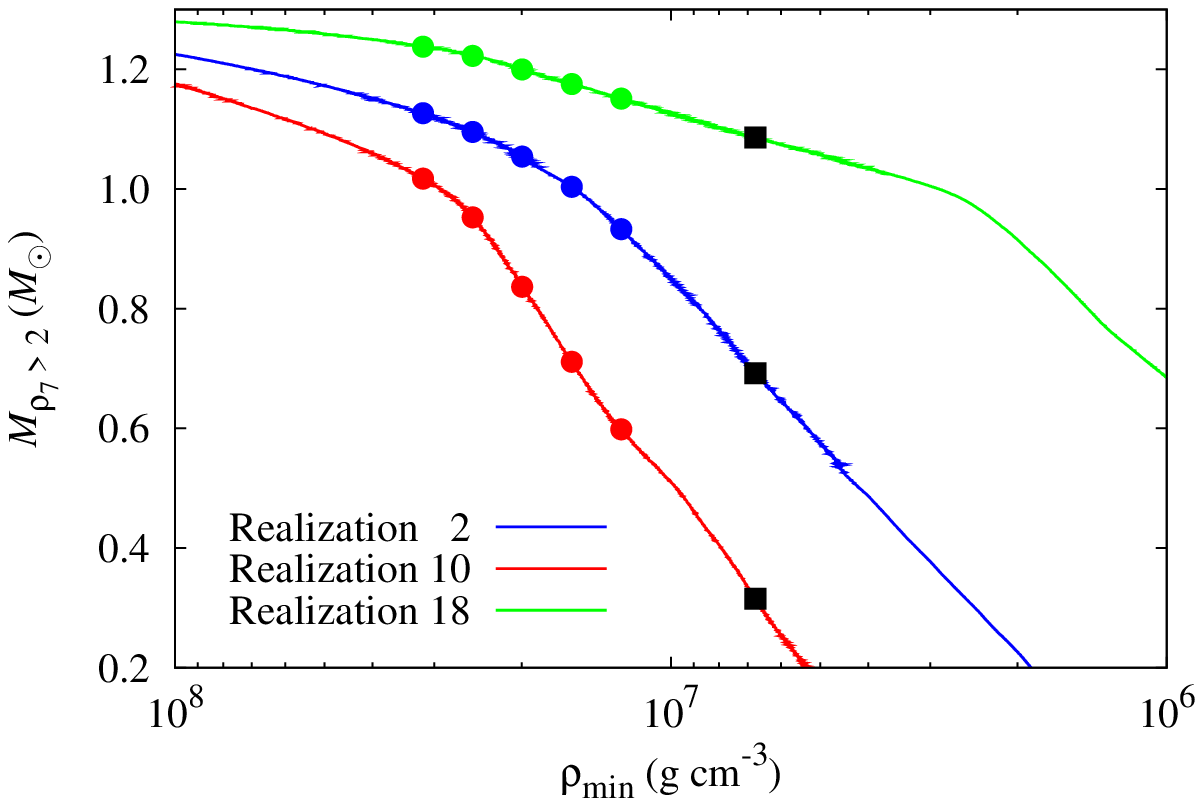}
\caption{Comparison of
the evolution of the white dwarf defined by the mass above the density
threshold ($M_{\rho_7>2}$) to the plume evolution defined by  the
minimum flame density ($\rho_{\rm min}$). The left panel presents the
evolution of $M_{\rho_7>2}$ (solid) and $\rho_{\rm min}$ (dotted) as a
function of simulation time for realizations 2 (blue), 10 (red), and 18
(green) showing the expansion rate of the WD and the plume rise-time
with DDT times (circles) associated with each DDT density emphasized.
$M_{\rho_7>2}$ is defined as the total mass with density greater than
$2\times 10^7$~g~cm$^{-3}$ and $\rho_{\rm min}$ is defined as the
minimum density at which the flame is burning material. The translation
into $M_{\rho_7>2}$ as a function of $\rho_{\rm min}$ is shown in the
right panel for comparison to $M_{\rm NSE}$ as a function of $\rho_{\rm
DDT}$ shown in \figref{fig:nsefit}. We also highlight the fiducial
DDT density as black squares. These plots emphasize the importance of
the rate of expansion of the WD and the plume rise-time as a function
of density in determining the relation between the NSE yield and DDT
density. Realization 2 produces NSE yields that are close to the average
while realizations 10 and 18 produce the lowest and highest yields,
respectively.
\label{fig:m2e7}
}
\end{center}
\end{figure*}

The evolution of the amount of material above a density threshold
($M_{\rho > \rho_{\rm thres}} = \int_{\rho_{\rm thres}} \rho dV$, where we
take $\rho_{\rm thres} = 2\times 10^7$~g~cm$^{-3}$) principally determines
the dependence of the NSE yield on DDT density. This is due largely to
the linear relationship between the NSE yield and $M_{\rho_7>2}(t=t_{\rm
DDT})$ shown in \figref{fig:Mrho}, where $t_{\rm DDT}$ is defined as the
time the flame first reaches the specified DDT density. 
This definition is consistent with our assumption that DDT conditions
are met on the tops of rising plumes.  As mentioned above, other DDT
locations are possible, such as the highly turbulent region underneath
plume ``caps'', which would result in a higher DDT density correlated
to our present definition via the density structure local to the rising
plume. As shown in \figref{fig:nse_evol}, the deflagration phase burns
only a relatively small fraction of the white dwarf and the majority of
material is burned during the detonation. Once a detonation has started,
the propagation speed of the burning wave is much greater than the
rate of expansion; therefore, the NSE yield is essentially independent
of the evolution of $M_{\rho_7>2}$ for $t > t_{\rm DDT}$. The number
of detonation points and their corresponding distribution in time and
location contributes to the variance of the relation between the NSE yield
and $M_{\rho_7>2}$. The relation between the NSE yield and DDT density
via $M_{\rho_7>2}$ also depends on the acceleration of the RT-unstable
plumes not being too great near $t_{\rm DDT}$. For a constant plume
rise-rate near $t_{\rm DDT}$, there is a linear relationship between the
DDT density and $t_{\rm DDT}$. This relationship allows the evolution of
$M_{\rho_7>2}$ to translate directly into a dependence of the NSE yield
on DDT density. \figref{fig:m2e7} shows the evolution of $M_{\rho_7>2}$
and $\rho_{\rm min}$ as a function of simulation time for 3 different
realizations where $\rho_{\rm min}$ is the minimum density at which the
flame is burning material. The evolution of $\rho_{\rm min}$ shows the
linear relationship between the log of the DDT density and $t_{\rm DDT}$
where the filled circles represent the $t_{\rm DDT}$ times for each
of 5 transition densities and the filled squares show the fiducial
DDT density. For the times corresponding to the DDT conditions, the
evolution of $M_{\rho_7>2}$ is in a region where the rate of expansion
of the star becomes significant and $M_{\rho_7>2}$ begins to drop off
relatively quickly.

\begin{figure}[tbh]
\plotone{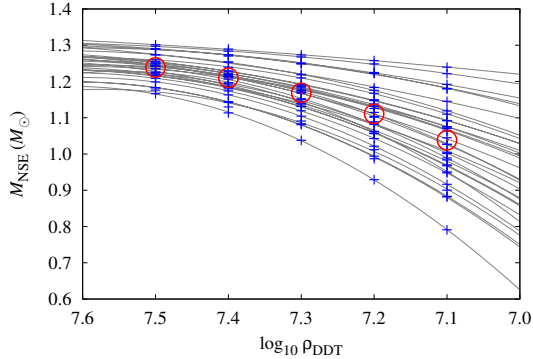}
\caption{\label{fig:nsefit}
Quadratic fits (grey lines) to the NSE yields for each realization
(blue crosses) and the average NSE yield at each transition density
(red circles). The qualitative similarity between this figure and the
right panel of \figref{fig:m2e7} is due to the near 1:1 correlation
between $M_{\rm NSE}$ and $M_{\rho_7>2}(t=t_{\rm DDT})$.
}
\end{figure}

We find, for each individual realization, the NSE yield depends
quadratically on transition density. \tabref{tab:nse} lists the
NSE yields for each realization at each of 5 transition densities
equidistant in log-space.  \tabref{tab:nsefit} shows the coefficients
used to fit the quadratic dependence of NSE yield on transition
density for 
\begin{equation}
\label{eq:nsefit}
M_{\rm NSE} (M_\odot) = a \left(\logtenrho \right)^2 + b \logtenrho + c
{\rm .}
\end{equation}
\figref{fig:nsefit} shows fits to the NSE yield for each realization
and the average yields at each transition density. The NSE yields for
individual realizations are plotted as blue crosses and the individual
curves showing the dependence of NSE yield on DDT density are shown in
grey. The average NSE yields at each DDT density are shown in red. At a
transition density of $10^{7.5}$~g~cm$^{-3}$, notice that the dependence
on DDT density has flattened out for almost all realizations and the
variance of the yields among all realizations is small.  This behavior
is due to the fact that the progenitor white dwarf star has a finite mass
of $1.37~M_\odot$. At the highest transition density, the expansion rate
of the star is relatively small, and the evolution of $M_{\rho_7>2}$
is relatively slowly declining as compared to lower DDT densities
(see \figref{fig:m2e7}).

\begin{table}
\caption{\label{tab:nse}NSE yields in $M_\odot$ for each realization at 
each transition density, $\logtenrho$.}
\begin{center}
\begin{tabular}{l||c|c|c|c|c}
Rel. & \multicolumn{5}{c}{$\logtenrho$} \\
\hline
\# & 7.1 & 7.2 & 7.3 & 7.4 & 7.5 \\
\hline
 1 & 0.989 & 1.082 & 1.152 & 1.207 & 1.236 \\
 2 & 0.970 & 1.056 & 1.131 & 1.184 & 1.215 \\
 3 & 0.948 & 1.043 & 1.119 & 1.174 & 1.209 \\
 4 & 0.951 & 1.062 & 1.144 & 1.197 & 1.226 \\
 5 & 1.027 & 1.101 & 1.168 & 1.214 & 1.242 \\
 6 & 1.074 & 1.138 & 1.185 & 1.219 & 1.248 \\
 7 & 1.076 & 1.132 & 1.179 & 1.215 & 1.243 \\
 8 & 1.110 & 1.167 & 1.210 & 1.239 & 1.260 \\
 9 & 0.917 & 1.022 & 1.105 & 1.163 & 1.199 \\
10 & 0.791 & 0.929 & 1.038 & 1.113 & 1.166 \\
11 & 1.092 & 1.147 & 1.189 & 1.229 & 1.255 \\
12 & 1.240 & 1.258 & 1.273 & 1.289 & 1.301 \\
13 & 0.882 & 0.987 & 1.083 & 1.145 & 1.183 \\
14 & 1.070 & 1.125 & 1.185 & 1.223 & 1.250 \\
15 & 1.119 & 1.176 & 1.218 & 1.254 & 1.274 \\
16 & 0.883 & 0.994 & 1.081 & 1.142 & 1.183 \\
17 & 1.026 & 1.115 & 1.171 & 1.213 & 1.241 \\
18 & 1.181 & 1.223 & 1.249 & 1.270 & 1.290 \\
19 & 0.967 & 1.057 & 1.132 & 1.185 & 1.220 \\
20 & 1.005 & 1.089 & 1.138 & 1.180 & 1.218 \\
21 & 1.222 & 1.248 & 1.268 & 1.284 & 1.297 \\
22 & 1.045 & 1.104 & 1.170 & 1.216 & 1.244 \\
23 & 1.001 & 1.085 & 1.150 & 1.187 & 1.223 \\
24 & 0.900 & 1.012 & 1.091 & 1.130 & 1.176 \\
25 & 0.989 & 1.057 & 1.139 & 1.193 & 1.226 \\
26 & 1.092 & 1.150 & 1.192 & 1.228 & 1.254 \\
27 & 1.192 & 1.222 & 1.251 & 1.274 & 1.290 \\
28 & 1.145 & 1.183 & 1.220 & 1.253 & 1.273 \\
29 & 1.063 & 1.131 & 1.175 & 1.210 & 1.235 \\
30 & 1.179 & 1.225 & 1.249 & 1.273 & 1.288 \\
\hline
$\bar{M}$ & 1.038 & 1.111 & 1.169 & 1.210 & 1.239 \\
$\sigma$  & 0.110 & 0.082 & 0.059 & 0.046 & 0.036 \\
$\sigma_{\bar{M}}$ & 0.020 & 0.015 & 0.011 & 0.008 & 0.007 \\
\hline
\end{tabular}
\end{center}
\end{table}

\begin{table}
\caption{\label{tab:nsefit}Coefficients used to fit the quadratic dependence 
of transition density on NSE yield using \eqref{eq:nsefit}. The NSE mass in 
units of $M_\odot$ is evalutaed at $\logtenrhoO$ using the coefficients.}
\begin{center}
\begin{tabular}{l||c|c|c||c}
\# & $a$ ($M_\odot$) & $b$ ($M_\odot$) & $c$ ($M_\odot$) & 
$M_{\rm NSE}(\logtenrhoO)$\\
\hline
  1 & -1.024 & 15.56 & -57.9 & 0.637 \\
  2 & -0.933 & 14.24 & -53.1 & 0.633 \\
  3 & -1.010 & 15.40 & -57.5 & 0.589 \\
  4 & -1.388 & 20.95 & -77.8 & 0.515 \\
  5 & -0.815 & 12.43 & -46.2 & 0.732 \\
  6 & -0.595 & 9.12  & -33.7 & 0.852 \\
  7 & -0.480 & 7.42  & -27.4 & 0.876 \\
  8 & -0.619 & 9.41  & -34.5 & 0.898 \\
  9 & -1.164 & 17.70 & -66.1 & 0.515 \\
 10 & -1.454 & 22.17 & -83.3 & 0.279 \\
 11 & -0.442 & 6.87  & -25.4 & 0.902 \\
 12 & -0.083 & 1.36  &  -4.3 & 1.183 \\
 13 & -1.212 & 18.45 & -69.0 & 0.456 \\
 14 & -0.557 & 8.58  & -31.8 & 0.844 \\
 15 & -0.575 & 8.78  & -32.2 & 0.910 \\
 16 & -1.173 & 17.88 & -66.9 & 0.470 \\
 17 & -0.963 & 14.58 & -54.0 & 0.712 \\
 18 & -0.348 & 5.34  & -19.2 & 1.048 \\
 19 & -0.926 & 14.15 & -52.8 & 0.628 \\
 20 & -0.708 & 10.86 & -40.4 & 0.740 \\
 21 & -0.210 & 3.25  & -11.3 & 1.134 \\
 22 & -0.583 & 9.03  & -33.6 & 0.798 \\
 23 & -1.093 & 16.47 & -60.9 & 0.696 \\
 24 & -1.219 & 18.46 & -68.7 & 0.502 \\
 25 & -0.782 & 12.04 & -45.1 & 0.671 \\
 26 & -0.490 & 7.56  & -27.9 & 0.896 \\
 27 & -0.248 & 3.86  & -13.7 & 1.080 \\
 28 & -0.276 & 4.36  & -15.9 & 1.006 \\
 29 & -0.685 & 10.42 & -38.4 & 0.827 \\
 30 & -0.429 & 6.54  & -23.6 & 1.032 \\
\hline
\end{tabular}
\end{center}
\end{table}

\begin{figure}[tbh]
\plotone{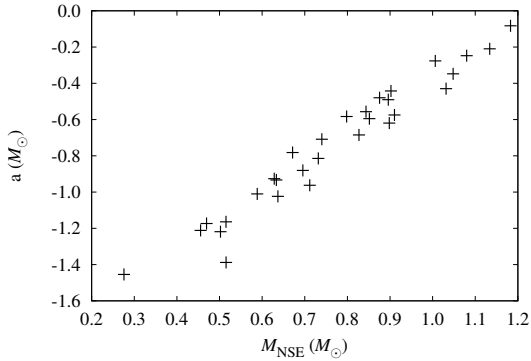}
\caption{\label{fig:nsecurve}
The correlation between the NSE yield at the fiducial DDT density
$\logtenrhoO=6.83$ and the fitting parameter $a$ from \eqref{eq:nsefit}.
A lower yield indicates a stronger curvature with DDT density.
}
\end{figure}

The curvature of the NSE yield dependence on DDT density, $a$, is well 
correlated with the NSE yield at a given DDT density. \figref{fig:nsecurve}
shows this correlation for $\logtenrhoO=6.83$; however, a correlation exists
for all DDT densities as may be discernible from \figref{fig:nsefit} noting 
that most black lines representing individual realizations do not cross.
The lower the NSE yield for a given realization, the stronger the dependence
on transition density. This result is likely due to realizations with lower
yields having multiple competing plumes that release more energy allowing
the star to expand more rapidly leading to a stronger dependence on the 
transition density.

\begin{figure}[tbh]
\plotone{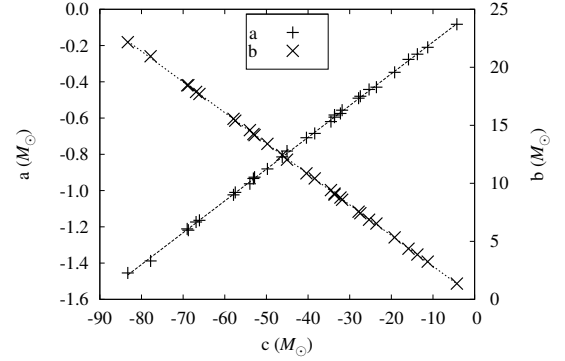}
\caption{\label{fig:nsefitcor}
The correlation between the NSE yield fitting parameter, $c$,
with the fitting parameters $a$ (left axis) and $b$ (right axis) from
\tabref{tab:nsefit} for each realization. The fitting parameters are defined
by \eqref{eq:nsefit} and the relations between the fitting parameters are
given by Equations~(\ref{eq:a}) and~(\ref{eq:b}). 
}
\end{figure}

After fitting the dependence on DDT density for each realization, we find an
interesting correlation between the fitting parameters from \eqref{eq:nsefit}. 
\figref{fig:nsefitcor} shows the correlation between the fitting parameters
$a$, $b$, and $c$ with $c$ on the horizontal axis.  The tight correlation 
between these parameters indicates that a single parameter describes the 
dependence on DDT density for a given realization.  The fitting parameters 
$a$ and $b$ can be expressed as a function of $c$:
\begin{eqnarray}
a &=& \alpha c + \beta \label{eq:a}\\
b &=& \delta c + \gamma \label{eq:b} {\rm ,}
\end{eqnarray}
where $\alpha = 1.754 \pm 0.003 \times 10^{-2}$, $\beta = -0.005 \pm 0.002$,
$\delta = -0.2646 \pm 0.0005$, and $\gamma = 0.21 \pm 0.02$. These parameters
were calculated using a Least Squares method and the associated errors were
calculated from the corresponding covariance matrix.

\begin{figure}[tbh]
\plotone{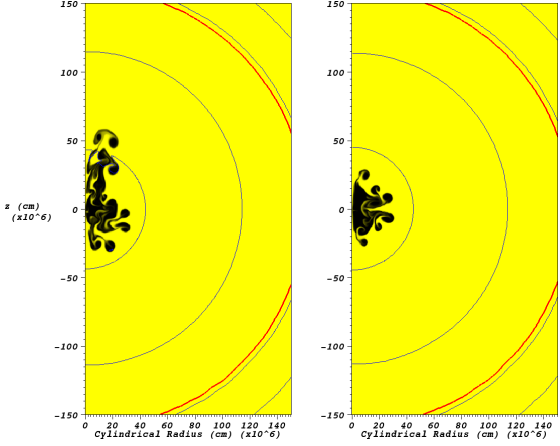}
\caption{\label{fig:plumecmp}Snapshots of realization 18 (left) and 
realization 10 (right) at a simulation time of 0.4 seconds.  Realization 18
produced the highest yield and shows the development of a single dominant plume
while realization 10 had the lowest yield and shows all plumes developing at 
about the same rate.  Shown in color are fuel and burning products:
unburned C, O, Ne (yellow) and Fe-group (NSE, black). Density in g~cm$^{-3}$
is indicated by contours (blue) logarithmically spaced at integer powers of 10
as well as the DDT density of $1.26\times10^7$~g~cm$^{-3}$ (red).
}
\end{figure}

Each realization has a different random initial 
perturbation of the central ignition condition that sets the plume 
morphology. Dominant single plumes tend to allow the star to expand less
prior to reaching the conditions for a detonation, while multiple competing 
plumes tend to release more energy during the deflagration phase, allowing 
more expansion prior to reaching the conditions for a DDT. Shown in 
\figref{fig:plumecmp} are the Rayleigh-Taylor unstable plumes for the 
realizations with the highest and lowest yields a few tenths of seconds 
into the deflagration. Unfortunately, no strong correlation exists between 
the properties of the initial flame surface, such as mass enclosed,
surface to volume ratio, or amount of power in the perturbation, for a given
realization and the single fitting parameter $c$.  This lack of a correlation
implies there is no way to tell whether a particular initial condition will
seed a single dominant plume or multiple competing plumes.

Physically, we might expect two competing effects that influence the NSE
yield: plume morphology (rise time) and the rate of expansion. In our 2D 
simulations these two effects appear to be correlated as seen in 
\figref{fig:m2e7}. Our results indicate 
that plumes near the symmetry axis tend to rise faster than plumes rising near 
the equator and we attribute this result to the fact that our simulations are 
two-dimensional. A plume that develops near the symmetry axis naturally becomes 
dominant in 2D and determines the DDT time and thus the NSE yield. In addition,
a plume near the symmetry axis represents less volume than a plume of similar
size near the equator. This indicates that energy is being deposited into a
smaller volume allowing for a lesser rate of expansion. This explains why the 
overall rate of expansion of the star is correlated to the rise time of the 
first plume to reach DDT conditions.

In order to evaluate the dependence on DDT density without this unphysical 
correlation, we need to perform 3D simulations. A suite of 3D simulations 
will likely create a two-parameter family (plume rise and expansion rates) 
of solutions which describe the yield as a function DDT density. Additionally,
we expect 3D plumes to behave similarly to 2D on-axis plumes implying a
faster plume rise time.

As a result of the tight correlation for $a$ and $b$ and the determination of
Equations~(\ref{eq:a},\ref{eq:b}), the average relation is characterized by 
just the average $c$---or equivalently, the average NSE yield at a specified 
$\rho_{\rm DDT}$.
We evaluate the average and standard deviation of the fitting parameter $c$
as well as the statistical properties of the NSE yield at the fiducial
transition density of $10^{6.83}$~g~cm$^{-3}$ shown in \tabref{tab:statfit}.

\begin{table}
\caption{\label{tab:statfit}Statistical properties of the fitting
parameter $c$ and the NSE yield at 
$\rho_{\rm DDT} = 6.76 \times 10^6$~g~cm$^{-3}$.
}
\begin{center}
\begin{tabular}{l||c|c|c}
   & Mean & Std. Dev. & Std. Dev. of Mean \\
\hline
$c$ ($M_\odot$) & -42 & 21 & 4 \\
$M_{\rm NSE,0}$ ($M_\odot$) & 0.77 & 0.22 & 0.04 \\
\end{tabular}
\end{center}
\end{table}

\subsection{NSE Yield Dependence on $\Ne{22}$}
\label{sec:results_nse}

Recall that for the purposes of this study, we are neglecting the effects
of varying the core carbon abundance, the central ignition density, the WD
structure, etc. except for the indirect effect of $\XNe$, the neutron excess
produced during simmering, and the DDT density. In order to derive the NSE 
yield dependence on $\XNe$, we need to expand the derivative of $M_{\rm NSE}$
with respect to $\XNe$ which involves only a couple of terms given our
assumptions
\begin{equation}\label{eq:dep}
  \der{M_{\rm NSE}}{\XNe} = \pder{M_{\rm NSE}}{\logtenrho}
    \der{\logtenrho}{\XNe} + \pder{M_{\rm NSE}}{\XNe}{\rm .}
\end{equation}
\cite{townetal09} showed that $\pder{M_{\rm NSE}}{\XNe} = 0$.
In that study, they employed an estimate of the mass burned to NSE by
measuring the amount of mass above a density of $2 \times 10^7$~g~cm$^{-3}$.
That correlation is confirmed by our current study as shown in
\figref{fig:Mrho} where we plot the NSE yield of all realizations at each
transition density. The correlation between NSE yield and mass above $2
\times 10^7$~g~cm$^{-3}$ at $t_{\rm DDT}$ for a transition density of 
$1.26\times 10^7$~g~cm$^{-3}$, closest to the transition density used for that 
study, is highlighted. Given our assumptions and the result that the NSE yield
does not depend directly on $\XNep$~\citep{townetal09}, the NSE yield 
only depends directly on the DDT density. The affect of $\XNep$ on the
yield enters through the DDT density. Therefore, we will construct the 
functional dependence of $\logtenrho$ on $\XNep$.

The DDT density depends on $\XNe$ via the microphysics involved in
determining the conditions under which the flame transitions from a
deflagration to a detonation, which, as noted above,
is incompletely understood. We assume the transition occurs at a
particular density at which the flame enters the distributed burning
regime, but more stringent conditions for the DDT include dependences
on the turbulent cascade and the growth of a critical mass of fuel with
sufficiently strong turbulence~\citep{Panetal08,Woosetal09,Schmetal10}.
The dependence on composition of these models may be explored and
applied to the trends with DDT density presented in this study. Under
our assumptions, the flame enters the distributed burning regime when
the laminar flame width becomes of order the Gibson length ($l_{\rm G}$)
where
\begin{equation}
\label{eq:gibson}
l_{\rm G} = L \left( \frac{s_l}{u'} \right)^3 {\rm .}
\end{equation}
Here, $L$ is the length scale on which the strength of the turbulent
velocity ($u'$) is evaluated, and $s_l$ is the laminar flame
speed~\citep[see e.g.][]{Peters00}. By using the laminar flame speeds
and widths which depend on $\XNe$ from~\cite{chamulak+07}, we choose the
conditions for a detonation are met at $\XNe = 0.02$ at a particular
transition density ($\logtenrho$) when the Gibson length is equal to
the laminar flame width.  The strength of the turbulent velocity field
is calculated assuming a Kolmogorov turbulence cascade:
\begin{equation}
\label{eq:turb}
u' = s_l(\rho,\XNe=0.02) 
    \left( \frac{\delta_l(\rho,\XNe=0.02)}{L} \right)^{1/3} {\rm .}
\end{equation}
Using $u'$, we solve for the change in transition density, $\Delta\logtenrho$, 
by changing $\Ne{22}$ to $\XNe = 0.06$ and again setting the Gibson length 
equal to the laminar flame width where $L$ cancels out of the equation. This 
evaluation yields a change in transition density over a change in $\XNe$, or 
$\der{\logtenrho}{\XNe}$.

We use log-log fits to the flame speed and flame width as a function of 
density using the table generated by \cite{chamulak+07}. We use only densities
above $10^8$ g cm$^{-3}$ for $\XC = 0.3$ and above $5 \times 10^8$ for $\XC
= 0.5$. Within this parameter space, a power-law dependence of the flame speeds
and widths on the log of the density is well-defined and the algorithm used
to solve the flame characteristics is more stable at higher densities.
Our resulting expressions for the density dependence on flame speed, $s$, and 
flame width, $\delta_l$, at the carbon mass fraction used in this study 
($\XC = 0.4$) for $\XNe = 0.02, 0.06$ are given by 
\begin{equation}
\label{eq:srho}
\ln{s} = a \ln{\rho} + b
\end{equation}
with coefficients given in \tabref{tab:flamtab}.


\begin{table}
\caption{\label{tab:flamtab}Coefficients for log-log fits to
\eqref{eq:srho}. The subscript denotes whether fitting for the
laminar flame speed ($s$) or flame width ($\delta$). }
\begin{center}
\begin{tabular}{l||c|c}
& $\XNe = 0.02$ & $\XNe = 0.06$ \\
\hline
$a_s$ & 0.7942 & 0.7745 \\
$b_s$ & -12.735 & -12.121 \\
$a_\delta$ & -1.3550 & -1.3507 \\
$b_\delta$ & 19.17 & 18.88 \\
\end{tabular}
\end{center}
\end{table}

Our derived expression for the derivative of transition density as a function 
of $\XNe$ is given by
\begin{eqnarray}
\der{\logtenrho}{\XNe} &=& \frac{ b_{\delta,6} - b_{\delta,2} + 3 
                           \left( b_{s,2} - b_{s,6} \right) + 
                           \ln{\rho_{\rm DDT}} 
                           \left( 3 a_{s,2} - a_{\delta,2} \right) }{
                           \ln{10} \left( 3 a_{s,6} - a_{\delta,6} \right) 
                           \Delta \XNe } \nonumber \\
&&- \frac{\logtenrho}{\Delta \XNe}  \nonumber \\
\label{eq:drhodne}
&=& u \logtenrho + v {\rm ,}
\end{eqnarray}
where $u = 0.4315$ and $v = -6.301$. We solve this first-order differential
equation and express the DDT density as a function of $\XNep$.
\begin{eqnarray}
\logtenrho(\XNep) &=& \frac{v}{u} 
  \left( e^{u \left( \XNep - \XNeOp \right)} - 1 \right) 
\nonumber \\
 &&+  \logtenrhoO e^{ u \left( \XNep - \XNeOp \right)} {\rm ,}
\label{eq:rhone}
\end{eqnarray}
where $\XNeOp$ is the parameterized $\Ne{22}$ mass fraction chosen to
be associated with the fiducial transition density, $\logtenrhoO$. For all the
transition densities considered in this study, the interface between the core
composition and the composition of the outer layer is at a lower density. 
Therefore, the relevant $\Ne{22}$ content to consider is that of the core. 
Plugging in \eqref{eq:rhone} into \eqref{eq:nsefit}, we obtain the functional
dependence of $M_{\rm NSE}$ on $\XNep$, such that
\begin{equation}\label{eq:mnsene}
M_{\rm NSE} = M_{\rm NSE}\left( \logtenrho \left( \XNep \right) \right)
{\rm .}
\end{equation}
 
\begin{figure}[tbh]
\plotone{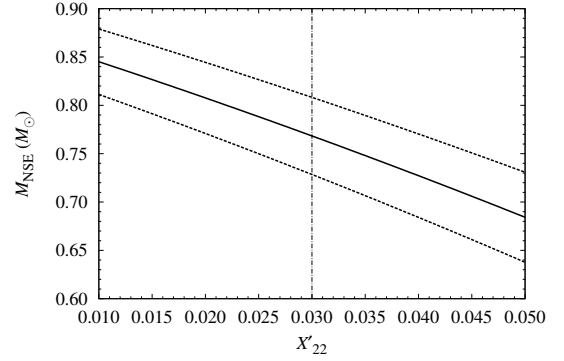}
\caption{\label{fig:mnsene}
The analytic solution to $M_{\rm NSE}$ as a function of 
parameterized $\XNep$ and using standard error propagation to
obtain the uncertainty based on the standard deviation of the mean 
(dashed lines). The vertical dot-dashed line indicates the parameter space 
in which this study was performed. These results were evaluated for the
fiducial transition density of $6.76 \times 10^6$~g~cm$^{-3}$ corresponding
to $\XNep=0.03$.
}
\end{figure}

\eqref{eq:mnsene} is evaluated and plotted in \figref{fig:mnsene} for the 
fiducial transition density at the $\Ne{22}$ mass fraction used in this study,
\begin{equation}
\label{eq:fiduc}
\logtenrho(\XNeOp=0.03) = \logtenrhoO = 6.83 {\rm .}
\end{equation}
We propagate the standard deviation of the mean evaluated at $\logtenrhoO$ for
a range of $\XNep$. This is calculated by considering the relation 
between the standard deviation of the mean of the NSE mass and the standard 
deviation of the mean of the fitting parameter, $c$, given by
\begin{equation}\label{eq:stdmrel}
\sigma_{\rm NSE}=\left(\alpha\left[\logtenrho\left(\XNep\right)\right]^2
                 + \delta
    \logtenrho \left( \XNep \right) + 1 \right) \sigma_c {\rm ,}
\end{equation}
where $\sigma_{\rm NSE}$ is the standard deviation of the mean of the NSE mass
and $\sigma_c$ is the standard deviation of the mean in the fitting parameter 
$c$. We evaluate $\sigma_c$ by inverting \eqref{eq:stdmrel} and solving for 
$\XNep = \XNeOp$. Plugging in this solution, the standard 
deviation of the mean of the NSE mass as a function of $\XNep$ becomes
\begin{equation}\label{eq:stdmnse}
\sigma_{\rm NSE} \left( \XNep \right) = 
 \frac{\alpha \left[\logtenrho \left(\XNep \right)\right]^2 + 
  \delta \logtenrho \left( \XNep \right) + 1}{\alpha
        \left[\logtenrhoO\right]^2 + 
   \delta \logtenrhoO + 1} \sigma_{{\rm NSE},0} {\rm .}
\end{equation}

\subsection{$\Ni{56}$ Yield Dependence on Metallicity}
\label{sec:results_m56}

Now that we have constructed the functional dependence of NSE yield on
$\Ne{22}$ we need to consider the dependence of the amount of $\Ni{56}$
synthesized in the explosion on metallicity through the $\Ne{22}$ content.
A fractional amount of the NSE material is radioactive $\Ni{56}$, which powers
the supernova light curve.  This fraction depends on $Y_e$ (and $\Ne{22}$) 
as described in \citet{timmes.brown.ea:variations}:
\begin{equation}
\label{eq:nseto56}
M_{56} = f_{56} M_{\rm NSE} {\rm .}
\end{equation}

In order to determine the dependence of $M_{56}$ on metallicity through the
DDT density, we must construct $f_{56}$ and explore any dependencies on DDT 
density and metallicity.
From our simulations, we calculate $M_{\rm NSE}$ directly and estimate $M_{56}$
from $\Ye$ choosing the non-$\Ni{56}$ NSE material to be 50/50 $\Fe{54}$ 
and $\Ni{58}$ by mass. Recall that $\XNep$ from \eqref{eq:x22param} contains
a term, $\Delta\XNe$, which is a parameterization of the change in $\Ye$ due to 
neutronization during the carbon simmering phase. However, not all of the neutron excess
evaluated at the $M_{\rm NSE}$ plateau time comes from $\Ne{22}$ or the 
carbon simmering products. Some change in $\Ye$ is due to weak 
reactions occuring during the explosion that are included in our burning 
model~\citep{Caldetal07,Townetal07}.

First we consider the amount of non-$\Ni{56}$ NSE material determined by $\XNep$
by equating the initial $\Ye$ to the $\Ye$ of material in NSE.
Using baryon and lepton conservation for NSE material, we describe
the electron fraction by contributions from $\Fe{54}$, $\Ni{58}$ , and $\Ni{56}$,
\begin{equation}
\label{eq:ye_nse}
\Ye = \frac{26}{54}\left(\frac{1}{2}f_{{\rm non}-56}\right) + 
\frac{28}{58}\left(\frac{1}{2}f_{{\rm non}-56}\right) + 
\frac{28}{56}\left(1-f_{{\rm non}-56}\right) {\rm ,}
\end{equation}
where $f_{{\rm non}-56}$ is the mass fraction of non-$\Ni{56}$ NSE material. 
For the following evaluation, we
approximate the composition to be that of the core because most of the NSE 
material is within the core.
We write the initial $\Ye$ as
\begin{equation}
\label{eq:yei}
Y_{e,i} = \frac{10}{22}\XNep + \frac{1}{2}\left(1-\XNep\right) 
{\rm .}
\end{equation}
We can then solve for $f_{{\rm non}-56}$ by equating (\ref{eq:ye_nse}) and
(\ref{eq:yei}) and add a term $X_n$ to represent the additional 
contribution due to neutronization from weak reactions occuring during the explosion 
\begin{equation}
\label{eq:non56}
f_{{\rm non}-56} = \frac{783}{308}\XNep + X_n {\rm .}
\end{equation}
Then the $\Ni{56}$ fraction of material in NSE is
\begin{equation}
\label{eq:tbt03}
f_{56} = 1 - \frac{783}{308}\XNep - X_n {\rm .}
\end{equation}
We note that the rate of weak reactions occuring during the 
explosion may depend on the initial composition as well. Accordingly,
$X_n$ may have a dependence on $\XNep$. We also expect $X_n$ to vary as 
a function of the transition density because a higher transition density 
will have less time for weak reactions to occur. 

\begin{figure}[tbh]
\plotone{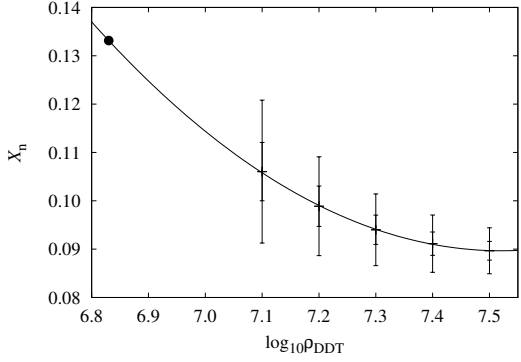}
\caption{\label{fig:xn} The fraction of non-$\Ni{56}$ NSE material due to
weak processes ($X_{n}$) is calculated from inverting \eqref{eq:tbt03} 
and using the $f_{56} = M_{56} / M_{\rm NSE}$ averaged over all
realizations at each DDT density. The standard deviation of the sample (outer)
and mean (inner) are shown as error bars at each DDT density. A quadratic best
fit line is calculated using the standard deviation of the sample. The
fiducial DDT density is shown as a black circle where $X_n = 0.133$.
}
\end{figure}

We construct a statistical sample of $f_{56}$ using the ratio of the $\Ni{56}$
and NSE yields produced in the simulations.
We calculate the dependence of $X_n$ on transition density
using a least-squares method and the result is shown in  \figref{fig:xn}.
Evaluating the partial derivative of $X_n$ at the fiducial transition density,
we find a shallow dependence:
\begin{equation}
\label{eq:Xnddt}
\left.\pder{X_n}{\logtenrho}\right|_{\logtenrho=\logtenrhoO} = -0.096 {\rm .}
\end{equation}

We want to know the dependence of $M_{56}$ on $\XNe$. Since $f_{56}$ depends
on the neutronization from weak reactions during the explosion, we must
consider any dependencies of this neutronization on DDT density or 
$\XNep$. We show that the dependence of $f_{56}$ on $\XNep$ via
the effect of $\XNep$ on the in-situ neutronization is much weaker than 
the direct dependence due to lepton number conservation. Similar to 
\eqref{eq:nseto56}, we write
\begin{equation}
\label{eq:dm56}
\der{M_{56}}{\XNe} = \pder{M_{56}}{M_{\rm NSE}} \der{M_{\rm NSE}}{\XNe} +
                     \pder{M_{56}}{\XNe} {\rm .}
\end{equation}
Using \eqref{eq:nseto56} and expanding the partial derivative of $M_{56}$ on
$\XNe$, we obtain
\begin{equation}
\label{eq:dm56dx22}
\der{M_{56}}{\XNe} = f_{56} \der{M_{\rm NSE}}{\XNe} +
       M_{\rm NSE}\der{f_{56}}{\XNe} {\rm .}
\end{equation}
Taking the derivative of \eqref{eq:tbt03} and expanding on $\XNe$, we obtain
\begin{equation}
\label{eq:dx56dx22}
\der{f_{56}}{\XNe} = \pder{f_{56}}{\XNe} - \der{X_n}{\XNe}
{\rm .}
\end{equation}
Now we wish to evaluate whether $\der{X_n}{\XNe}$ is an important
contribution to the overall evaluation of the $\Ni{56}$ mass. Expanding
this term yields
\begin{equation}
\label{eq:dxndx22}
\der{X_n}{\XNe} = \pder{X_n}{\logtenrho} \der{\logtenrho}{\XNe} + 
   \pder{X_n}{\XNe} {\rm .}
\end{equation}
Referring to \cite{townetal09}, we can estimate $\pder{X_n}{\XNe}$ by 
calculating the average ratio of $M_{56}$ to $M_{\rm NSE}$ at $\XNe=0, 0.02$
for the first 5 realizations whose detonation phases were simulated. The 
result is $\pder{X_n}{\XNe} \sim -0.2$. The first term in \eqref{eq:dxndx22}
can be evaluated from multiplying \eqref{eq:Xnddt} and \eqref{eq:drhodne}
for $\logtenrho$ in the range 7.0--7.5. For the fiducial transition density
($\logtenrhoO$), we find
\begin{equation}
\label{eq:dxndx22_eval}
\der{X_n}{\XNe} = \left( -0.126 \right) \left( -3.35 \right) - 0.2 \sim 0.2
{\rm .}
\end{equation}
Comparing to $\pder{f_{56}}{\XNe} \simeq -2.5$, we find that the magnitude
of $\der{X_n}{\XNe}$ is much smaller and is unimportant for our evaluation
of the dependence of the $\Ni{56}$ yield on $\XNe$. Therefore, we 
can ignore this term in the expansion of $\der{f_{56}}{\XNe}$ in
\eqref{eq:dx56dx22} and let $\der{X_n}{\XNe} \sim 0$. The full derivative
of $f_{56}$ with respect to $\XNe$ can now be written as a partial derivative
such that
\begin{equation}
\label{eq:dm56_3}
\der{M_{56}}{\XNe} = f_{56} \der{M_{\rm NSE}}{\XNe} + 
                     M_{\rm NSE} \pder{f_{56}}{\XNe} {\rm .}
\end{equation}

\begin{figure}[tbh]
\plotone{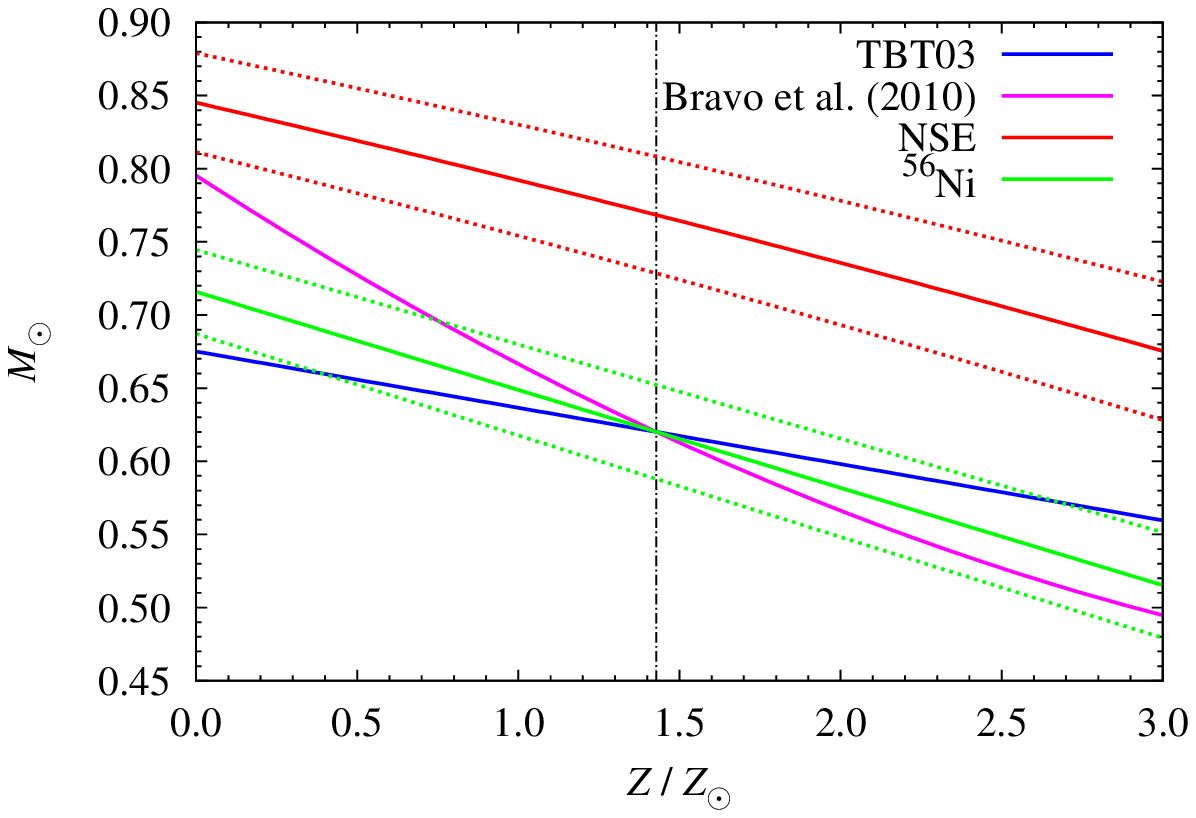}
\caption{\label{fig:sum}
The solution of $M_{\rm NSE}$ (red) and $M_{56}$ (green) computed as a
function of metallicity as compared to the $\Ni{56}$ relations 
from \citet{timmes.brown.ea:variations} (blue) and \citet{bravo10} (magenta) 
normalized to the average $\Ni{56}$ yield from our simulations. The dashed
lines show the propagated standard deviation of the mean. The vertical 
dot-dashed line indicates the parameter space in which this study was 
performed. These results were evaluated with a fiducial transition density of
$6.76 \times 10^6$~g~cm$^{-3}$ at $Z= 1.4 Z_\odot$.
}
\end{figure}

While we approximate $X_n$ as constant, we choose to evaluate it at 
$\logtenrhoO$ using the best fit curve from \figref{fig:xn} obtaining 
$X_n = 0.133$. Now we can relate the metallicity 
to $\Ne{22}$ since $\XNe$ traces metallicity~\citep{timmes.brown.ea:variations}.
Substituting $\XNe = 0.014 (Z / Z_\odot)$ in \eqref{eq:x22param}, we obtain
\begin{equation}
\label{eq:nemetal}
\XNep = 0.014 \left( \frac{Z}{Z_\odot} \right) + \Delta\XNe(\Delta\Ye) 
{\rm ,}
\end{equation}
where $\Delta\XNe=0.01$ for this study and $\XNep$ is our parametrization
of the actual $\Ne{22}$ mass fraction, $\XNe$. The $\Ni{56}$ yield as a
function of $\XNep$ is calculated by multiplying \eqref{eq:tbt03} by
\eqref{eq:mnsene} and using $X_n = 0.133$. The $\Ni{56}$ yield is plotted
in \figref{fig:sum} using \eqref{eq:nemetal} to relate $\XNep$ to 
$Z / Z_\odot$.

\section{Conclusions and Future Work}
\label{sec:conclusions}

We have analyzed the influence of the DDT density on the total $\Ni{56}$ 
synthesized during a thermonuclear supernovae.  We determined that the 
dependence on DDT density is quadratic in nature, but this dependence can be
described by a single parameter. 
We estimated the dependence of the supernova brightness ($\Ni{56}$ yield)
on metallicity by assuming a DDT occurs when the flame enters the distributed
burning regime and extrapolating the laminar flame speeds and widths down to
low densities. We find
\begin{equation}
\label{eq:m56z}
\left.\der{M_{56}}{(Z/Z_\odot)}\right|_{Z=Z_\odot} = -0.067 \pm 0.004 M_\odot
{\rm ,}
\end{equation}
which is slightly steeper than \citet{timmes.brown.ea:variations} as seen
in \figref{fig:sum}. The uncertainty was calculated using the standard
deviation of the mean of the fitting parameter $c$---or equivalently,
the standard deviation of the mean of the NSE yield at a particular DDT
density. The uncertainty in the assumptions about the normalization of
the transition density as a function of $\XNep$ was not considered for
the purpose of evaluating the uncertainty in the derivative.  We find
the effect of metallicity on DDT density influences the production of
NSE material; however, the ratio of the $\Ni{56}$ yield to the overall
NSE yield does not change as significantly and remains similar to the
relation estimated from approximate lepton number conservation.
We also find that the scatter in supernova brightness 
increases with decreasing transition density. 


The very recent work of \citet{bravo10} on metallicity as a source of 
dispersion in the luminosity-width relationship of bolometric light
curves addresses many of the same issues as our study and warrants
discussion. In particular, \citet{bravo10} derive a metallicity dependence
on the DDT density similarly to our study. However, the DDT density
in their 1D simulations is very different from the DDT density in ours.
The principal difference between the work described in this manuscript 
and that of \citet{bravo10} follows from the use of multi-dimensional
simulations. While 3D simulations are required to correctly capture the 
effects of fluid dynamics and the Rayleigh-Taylor instability, 2D 
simulations incorporate these effects and relax the assumption of 
symmetry. Breaking the symmetry and assuming that a DDT initiates as 
a rising plume approaches $\rho_{\rm DDT}$ produces an expansion
history very different from what would be observed in 1D simulations.
In fact, using 1D simulations implies that $\rho_{\rm DDT}$ plays an 
unphysical role. The real physical description of the DDT in a type Ia 
supernova will depend on flame-turbulence interactions that will 
themselves depend on multi-dimensional effects in the flow. 1D models may
be able to capture these effects, but such models must be motivated by 
more physical multi-dimensional studies.

In addition, our statistical framework with realizations from randomized 
initial conditions allows calculation of the scatter inherent in the 
multi-dimensional models. The standard deviations calculated for our 
models averaged over a set of realizations demonstrate that there can
be considerable variation following from the randomized initial 
conditions. These variations follow from the different amounts of 
expansion occurring during the deflagration phase that follow from 
the different rising plume morphologies. In 1D models, even with the 
progenitor metallicity determining $\rho_{\rm DDT}$, there is a 
one-to-one correspondence between expansion and $\rho_{\rm DDT}$ for 
a given progenitor. 

\citet{bravo10} explored two scenarios, a linear and a 
non-linear dependence of the $\Ni{56}$ yield on $Z$, and report excellent 
agreement between the non-linear scenario and observations reported by
\citet{gallagheretal+08} with the caveat that they used 1D models to arrive
at their conclusion. Without comparing results involving the metallicity
dependence on $\rho_{\rm DDT}$, \citet{bravo10} find a stronger dependence
of the $\Ni{56}$ synthesized from incomplete Si-burning on $Z$ than $\Ni{56}$
coming from NSE material. We find a shallower dependence of the $\Ni{56}$ 
yield on $Z$ than \citet{bravo10} in part due to our assumption that 
$\Ni{56}$ is synthesized as a fraction of NSE material; however, comparing 
their stratified models to the \citet{timmes.brown.ea:variations} relation 
results in only a $\sim 20\%$ difference in the dependence of $\Ni{56}$ yield 
on $Z$. This effect is compounded in their ``non-linear'' scenario, but again,
we must emphasize that $\rho_{\rm DDT}$ plays a completely unphysical role 
in their simulations and the degree to which this effect is actually enhanced
is unknown.

The light curve width calculations and subsequent population
synthesis performed by \citet{bravo10} after finding a dependence
of the $\Ni{56}$ yield on metallicity is useful in estimating a
metallicity dependence in the hubble residual. For this purpose,
we compare the results of the non-linear scenario of \citet{bravo10}
to our study along with the expected dependence due to lepton number
conservation~\citep{timmes.brown.ea:variations} in \figref{fig:sum}.
\citet{bravo10} find a $\sim 30\%$ steeper dependence of the $\Ni{56}$
yield on $Z/Z_\odot$ evaluated at $Z= 1 Z_\odot$ than our result in
\eqref{eq:m56z}.  In addition, they find that for high $Z$ the $\Ni{56}$
yield tends to flatten out becoming more similar to our results. For
a particular subrange of metallicites, our results are very similar
indicating that by performing the same analysis as \citet{bravo10}
our results should also agree with the metallicity dependence found by
\citet{gallagheretal+08}.

This similarity depends on the choice of mean $\Ni{56}$ yield as both 
\citet{timmes.brown.ea:variations} and \citet{bravo10} relations are 
proportional to this quantity. This implies that the steepness of the 
relations are affected by the choice of mean $\Ni{56}$ yield. Our results 
are not sensitive to this choice. Choosing a higher fiducial DDT density 
of $10^7$~g~cm$^{-3}$ that results in a higher mean $\Ni{56}$ yield of 
$0.77~M_\odot$ increases \eqref{eq:m56z} by only $0.005~M_\odot$. We
see less dependence on the mean $\Ni{56}$ yield because, although we
have a similar dependence on $\XNe$ due to lepton number conservation,
the dependence of the yield on $\logtenrho$ is stronger at lower fiducial
$\logtenrhoO$, as shown in previous sections. This effect is not captured
in a simple proportionality relation like that quoted by \citet{bravo10},
even if this effect is present in their calculations.

It has been discussed whether the source of scatter in peak brightness
itself can be attributed to metallicity~\citep{timmes.brown.ea:variations,
howelletal+09,bravo10}.  We submit that multi-dimensional effects
following from fluid instabilities during the deflagration phase
leading to varying amounts of expansion provide scatter consistent with
observations. The influence of metallicity on the DDT density affects
the duration of the deflagration which will secondarily influence the
magnitude of the scatter. Our results show that the primary parameter is
the degree of (pre-)expansion before DDT---which determines the amount of
mass at high density. This is, in turn, controlled by both the expansion
rate and the plume rise time. These are also expected to be the basic
ingredients in reality.

While our study is consistent with the expected theoretical brightness 
trend with metallicity~\citep{timmes.brown.ea:variations}, observations
to date have not been able to confirm this prediction~\citep{GallGarnetal05,
gallagheretal+08,howelletal+09}. The progenitor age is difficult to decouple
from metallicity given the mass-metallicity relationship within
galaxy types~\citep{gallagheretal+08}. Additionally, the dependence of
mean brightness of SNe~Ia on the age of the parent stellar population appears
to be much stronger than any dependence on metallicity~\citep{gallagheretal+08,
howelletal+09}. Seemingly, the only way to observe a dependence on metallicity
is to constrain the mean stellar age by selecting galaxies of the same type.
This approach was used by \citet{gallagheretal+08}, but the difficulties
in accurately measuring metallicities for the parent stellar population have
only constrained the magnitude of the effect on mean brightness, but thus 
far have not proved that a metallicity effect exists.

In order to determine the dependence of supernova brightness ($\Ni{56}$
yield) on metallicity, a better understanding of how the transition
density is affected by changes in the $\Ne{22}$ content is needed.
Currently, we have extrapolated data from \cite{chamulak+07} down to the
range of expected transition densities using flame data from densities
above $10^8$~g~cm$^{-3}$.  The trend observed by \citet{chamulak+07} that
increasing $\XNe$ increases the laminar flame speed is valid for densities
above $10^8$~g~cm$^{-3}$; however, it is unclear if this trend will hold
for densities below $10^7$~g~cm$^{-3}$. 
 Direct numerical simulations of the flame are needed for these
lower densities to determine whether this trend still holds.  However, in
this parameter space, the laminar flame is extremely slow requiring a
low-Mach-number treatment to model the flow.
Regardless of the
choice of model for the mechanism that produces a spontaneous detonation,
a necessary condition is thought to be distributed burning. Properties
of the laminar flame are necessary to estimate when burning becomes
distributed. Future studies exploring other DDT mechanisms must also 
determine their compositional dependencies.
So far, we have only explored the effect of varying the $\Ne{22}$ content
directly~\citep{townetal09} and indirectly through the DDT density. Other
effects exist that we have yet to study such as the carbon composition
and flame ignition density (set by the average progenitor age). These
properties may be influenced by metallicity such that the net affect on
mean brightness is negligible. Additionally, if our assumptions about
the dependence of DDT density on metallicity are incorrect and further
studies indicate the opposite trend, the metallicity effect on DDT
density could negate the effect due to lepton number conservation.

This study stresses the importance of multi-dimensional effects, but has
provided evidence of the limitations of 2D simulations of supernovae. In
the immediate future, we plan to perform 3D simulations for better
realism.  In reality, we might expect a two-parameter family in which the
morphology of the dominant Rayleigh-Taylor plume is independent of the
rate of expansion; however, for our 2D simulations, these two effects
appear to be correlated by the choice of cylindrical geometry. A plume
developing along the symmetry axis is like 3D RT compared to one near
the equator, which is more like 2D RT, and the latter is weaker (slower
to develop). In addition to developing faster, plumes near the symmetry
axis represent less volume and, therefore, expand the star less. A fast
rising plume combined with a slow rate of expansion indicates there
will be a shallow dependence on DDT density (since the plume will move
through the various DDT densities faster and with less expansion).
This result may also depend on our choice that DDT
conditions are met at the tops of rising plumes. For these reasons,
3D studies are necessary to ascertain whether there is a physically
motivated correlation between the dominant plume morphology and the rate of
expansion and whether this correlation depends on the choice
of location of the DDT. In transitioning to 3D simulations, we
expect the growth of many more plumes; however, the rate of expansion
will likely not exceed that found in 2D. Because RT develops faster in
3D, the conditions for DDT will be met sooner leading to less overall
expansion of the star and more mass at high density at the first DDT
time. Therefore, we expect 3D results similar to the higher yielding 2D
realizations with a shallower dependence on DDT density.  In any case,
this study indicates that it is possible to determine the susceptibility
of a particular model to a change of DDT density (and, hence, a change in
the composition which alters the microphysics that set the DDT density).
Once the plume morphology has been established several tenths of seconds
into the simulated explosion, it should be possible to estimate the
total $\Ni{56}$ yield to fairly high accuracy given DDT conditions. However,
the treatment of turbulent flame properties may be important.


\acknowledgements

This work was supported by the Department of Energy through grants
DE-FG02-07ER41516, DE-FG02-08ER41570, and DE-FG02-08ER41565, and by NASA 
through grant NNX09AD19G.  ACC also acknowledges support from the 
Department of Energy under grant DE-FG02-87ER40317. DMT received support
from the Bart J. Bok fellowship at the University of Arizona for part of 
this work. The authors acknowledge 
the hospitality of the Kavli Institute for Theoretical Physics, which is 
supported by the NSF under grant PHY05-51164, during the programs ``Accretion 
and Explosion: the Astrophysics of Degenerate Stars'' and ``Stellar Death and
Supernovae.''  The software used in 
this work was in part developed by the DOE-supported ASC/Alliances Center 
for Astrophysical Thermonuclear Flashes at the University of Chicago. We 
thank Nathan Hearn for making his QuickFlash analysis tools publicly 
available at http://quickflash.sourceforge.net. We also
thank the anonymous referee for a careful reading of the manuscript and 
constructive comments that improved this work.
This work was
supported in part by the US Department of Energy, Office of Nuclear
Physics, under contract DE-AC02-06CH11357 and utilized resources at 
the New York Center for Computational Sciences at Stony Brook 
University/Brookhaven National Laboratory which is supported by the U.S. 
Department of Energy under Contract No. DE-AC02-98CH10886 and by the State of 
New York.  

\bibliography{master,timmes_master,townsley_master}

\end{document}